\documentclass[12pt, preprint]{aastex}

\slugcomment{To appear in the Astrophysical Journal}
\begin{document}

\newcommand{\be}{\begin{equation}}
\newcommand{\ee}{\end{equation}}

\def\xr#1{\parindent=0.0cm\hangindent=1cm\hangafter=1\indent#1\par}
\def\ea{et al.}

\def\snei{{\it NEI }}
\def\vsnei{{\it VNEI }}
\def\nei{{\it GNEI }}
\def\vnei{{\it VGNEI }}
\def\shock{{\it shock}}
\def\vshock{{\it vpshock}}
\def\nshock{{\it NPSHOCK }}
\def\vnshock{{\it VNPSHOCK }}
\def\sedov{{\it SEDOV }}
\def\vsedov{{\it VSEDOV }}
\def\srcut{{\it SRCUT }}
\def\vmekal{{\it VMEKAL }}
\def\mekal{{\it MEKAL }}
\def\power{{\it POWER }}

\def\xr#1{\parindent=0.0cm\hangindent=1cm\hangafter=1\indent#1\par}
\def\ea{\it et al.\rm}
\def\exo{{\sl EXOSAT}}
\def\ein{{\sl EINSTEIN}}
\def\gin{{\sl GINGA}}
\def\rosat{{\sl ROSAT}}
\def\asca{{\sl ASCA}}
\def\chandra{{\sl CHANDRA}}
\def\astroe{{\sl ASTRO-E2}}

\def\sest{{\sl SEST}}
\def\as{$^{\prime\prime}$ }
\def\am{$^{\prime}$}
\def\rxte{{\sl RXTE}}
\def\xmm{{\sl XMM-Newton}}
\def\cps{{ ct~s$^{\rm -1}$ }}
\def\ergpsec{{ergs~s$^{\rm -1}$}}
\def\ergpcps{{ergs~cm$^{\rm -2}$~s$^{\rm -1}$}}
\def\3c{3C\,397}
\def\g{G41.1--0.3}
\def\cxo{CXO J190741.2+070650}

\def\shock{{\it vpshock}}
\def\nei{{\it vnei }}

\def\sol{$\odot$}
\def\solpyr{{M$_{\odot}$ yr$^{-1}$}}
\def\ergpsec{{ergs s$^{-1}$}}

\title{\chandra\ Spatially Resolved Spectroscopic Study and Multi-Wavelength
Imaging of the Supernova Remnant \3c (\g)}

\author{S. Safi-Harb\altaffilmark{1,2,3}, G. Dubner\altaffilmark{4,5}, R. Petre\altaffilmark{6},
 S. S. Holt\altaffilmark{7},
 P. Durouchoux\altaffilmark{8}
}
\altaffiltext{1}{ Department of Physics and Astronomy, University of Manitoba, Winnipeg,
MB, R3T 2N2, Canada}
\altaffiltext{2}{NSERC University Faculty Award fellow; samar@physics.umanitoba.ca}
\altaffiltext{3}{Department of Physics, The George Washington University, Washington, D.C. 20052}
\altaffiltext{4} {IAFE, Institute of Astronomy and Space Physics, Buenos
Aires, Argentina}
\altaffiltext{5}{Member of the Carrera del Investigador Cient\'\i fico
of CONICET, Argentina}
\altaffiltext{6} {NASA's Goddard Space Flight Center, Greenbelt, Maryland, 20771}
\altaffiltext{7}{F. W. Olin College of Engineering, Needham, MA 02492}
\altaffiltext{8} {Centre d'Etudes Nucleaires, Service d'Astrophysique, Saclay, France}

\begin{abstract}
We present a \chandra\ observation of the supernova remnant (SNR) \3c\ (\g)
obtained with the Advanced CCD Imaging Spectrometer (ACIS-S).
Previous studies of this SNR have shown that the
remnant
harbors a central X-ray `hot spot'
suggestive of a compact object associated with \3c.
With the \chandra\ data , we can rule out the nature of the hot spot as a pulsar
or a pulsar wind nebula, and put an upper limit on the flux of a hidden compact
object of $F_x$ (0.5--10 keV)$\sim$6$\times$10$^{-13}$~erg~cm$^{-2}$~s$^{-1}$.
We found two point sources in the observed \chandra\ field. We argue that
none of them is associated with \3c; and that the hard source, \cxo\ which
is characterized by a heavily absorbed spectrum with a strong Fe-line, is
a newly discovered active galactic nucleus.
The \chandra\ image
reveals arcseconds-scale clumps and knots 
which are strongly correlated with the radio VLA image,
except for the X-ray hot spot.
Our \chandra\ spatially resolved spectroscopic study
shows that one-component models are inadequate,
and that at least two non-equilibrium
ionization thermal components are needed to fit the 
spectra of each selected region. The 
derived average spectral parameters are consistent with 
the previous global \asca\ fits  performed by Safi-Harb and
coworkers.  However, the hard component
requires a high abundance of Fe indicating the presence of hot Fe ejecta.
When comparing the eastern with the western lobe, we find that the
column density, the brightness, and the ionization timescales are generally higher 
for the western side. This result, combined with
our study of the \3c\ environs at millimeter wavelengths,
% of the $^{12}$CO and the $^{13}$CO line
%transitions
%using the MOPRA and SEST telescopes and
%the International Galactic Plane Survey,
indicate a denser medium to the west of the SNR.
Our multi-wavelength imaging and spectral study favors the scenario
where \3c\ is a $\sim$5,300-year old SNR expanding in a medium
with a marked density gradient,
and which is likely to be encountering a molecular cloud on the western side. 
We propose that \3c\ will evolve into a mixed-morphology SNR.
\end{abstract}


\section{Introduction}
X-ray observations of SNRs provide crucial information on the
intrinsic properties of supernova explosions, the distribution of ejecta,
the nature of their collapsed cores,
and the conditions of the interstellar medium (ISM).
The morphology and dynamics of SNRs are highly shaped by both the progenitor star and the ISM. 
Explosions of massive stars are expected to leave behind neutron stars
which can power synchrotron nebulae (also called plerions or pulsar wind nebulae, PWNs).
PWNs give the SNR a centrally-filled morphology dominated by non-thermal hard X-ray emission.
The supernova ejecta and the shocked ISM form a shell-like component emitting
thermal X-rays.  SNRs having both the shell-like and centrally-bright components
are referred to as `Composites'. Some Composite-type SNRs, however, display a centrally-bright component
dominated by thermal emission arising
from the swept-up ISM. Rho \& Petre (1998) refer to this latter class as `mixed-morphology' or thermal Composites,
to distinguish them from the plerionic Composites.
In the pre-\chandra\ era, many compact objects or PWNe have been missed in SNRs. Furthermore, resolving
the various components of Composites was hampered by the lack of arcsecond resolution.
This left the classification of many SNRs as uncertain. 

3C\,397 (G41.1-0.3) is one such example.
In the radio, it is
classified as a shell-type SNR,  based on its
 steep spectral index ($\alpha$ = 0.48, $S_{\nu}$$\sim$$\nu^{-\alpha}$) and shell-like morphology 
(Green 2004)\footnote{http://www.mrao.cam.ac.uk/surveys/snrs/}.
In X-rays, it has been classified as a Composite because of a central X-ray enhancement; however
whether it is thermal Composite remained highly uncertain (Rho \& Petre 1998).
High-resolution radio imaging of \3c\
(Becker, Markert, \& Donahue 1985; Anderson \& Rudnick 1993) indicates
that the remnant brightens on the side closer to the Galactic plane, and is highly
asymmetric.  It  has  the appearance of a shell edge-brightened in
parts, and lacks the spherical symmetry seen in the young
historical SNRs, such
as Cas~A and Tycho. 
 Kassim (1989) derives an integrated
spectral index $\alpha$ = 0.4, with a turnover at a frequency less than
100 MHz.
The discrepancy between the spectral indices in Kassim (1989) and
Green (2004) is most likely due to uncertainties in measuring the
total flux density of 3C~397 at centimeter wavelengths, and is attributed
to confusion with a nearby HII region and the Galactic background.
 Anderson \& Rudnick
(1993)
 investigate the variations of the spectral index across
the remnant, and find variations of the order of $\delta$$\alpha$
$\sim$ 0.2 ($\alpha$ $\sim$ 0.5--0.7).
The variations do not coincide with variations in the total
intensity.  
They suggest that interactions between the expanding SNR
and inhomogeneities in the surrounding medium play a major role in
determining the spatial variations of the index across the remnant.

Dyer \& Reynolds (1999), while finding a similar magnitude
of spectral index variations, did not confirm Anderson \& Rudnick's
detailed spatial results, suggesting that the variations are due to
image reconstruction problems or other difficulties.
They find that the remnant is unpolarized at 20 cm and has a mean fractional polarization of 1.5\% at 6 cm.
 The polarization peaks inside the remnant 
at a location not coincident with either a radio feature or
with the X-ray hot spot. 
Spectral index maps between 6 and 20 cm do not show any systematic differences associated with interior emission; and no pulsar-driven component (which would be characterized by high
polarization and a flat radio spectrum) was found.

\rosat \ observations of 3C\,397 with the PSPC (Rho 1995, Rho \& Petre
1998, Dyer and Reynolds 1999, Chen et al. 1999,
Safi-Harb et al. 2000) reveal 2\farcm5 $\times$ 4\farcm5 diffuse emission, with
central emission and an enhancement along the western edge.
Most intriguing in the \rosat\ image is a central `hot spot' suggestive of a putative compact object or a small plerion; however it is not correlated
with any radio enhancement.
The narrow bandpass of \rosat \ did  not allow for an accurate determination of its X-ray emission
mechanism. 

The combined \rosat, \asca \ and \rxte \ study of the SNR
(Safi-Harb et al. 2000)
showed that the overall SNR spectrum is heavily absorbed
($N_{\rm H}$=3.1$^{+0.2}_{-0.3}$$\times$10$^{22}$ cm$^{-2}$), complex, 
and dominated by thermal emission from Mg, Si, S, Ar, and Fe.
A two-component thermal model provided an adequate fit: 
the soft component was characterized by a
low temperature ($kT$$\sim$0.2 keV) and a large ionization timescale ($\sim$6$\times$10$^{12}$~cm$^{-3}$ s),
and the hard component required to account for the Fe-K emission line was characterized by a
high temperature ($kT$$\sim$2~keV) and much lower ionization timescale ($\sim$6$\times$10$^{10}$~cm$^{-3}$ s).
The X-ray spectrum was discussed in the light of two scenarios:  a young
 ejecta-dominated remnant of a core-collapse SN, and a middle-aged SNR 
expanding in a dense ISM.  In the first scenario, the hot component
arises from the SNR shell, and the soft component from an
ejecta-dominated component.  3C\,397 would be then $\sim$2,000-year old,
young but intermediate in dynamical age between the
young historical shells (like Tycho or Kepler), and those that are well
into the Sedov phase of evolution (like Vela).  In the second scenario,
the soft component represents the blast wave propagating in a dense
medium, and the hard component is associated with hot gas encountering
a fast shock, or arising from thermal conduction.  In this latter
scenario, the SNR would be $\sim$5,300-year old, and in transition
into the radiative phase.
No pulsations were found associated with the central hot spot.
While the spectrum of the X-ray spot could not be resolved with \asca, 
a hardness ratio image did not reveal the hard X-ray emission 
 that would be expected from a PWN.

\3c\ has no optical counterpart and is not observed in the UV, probably because
of the high interstellar absorption towards the remnant which lies
in the Galactic plane.
Neutral hydrogen absorption measurements indicate that 
\3c\ is located between
6.4 kpc and 12.8 kpc (Caswell \ea \ 1975).
The HII region G41.1-0.2, lying $\sim$ 7\am \ west of the SNR
is likely a foreground object,
and was located between 3.6 kpc and 9.3 kpc (Cersosimo \& Magnani 1990).
At a distance of 10 kpc, the linear size of the radio `shell' would be
7$\times$13 pc$^2$.

\chandra\ observations were carried out to 1) study the nature of the `hot spot' and search for the compact stellar remnant,  2) resolve the soft and hard components,
 and 3) address the intriguing morphology and classification of this SNR.
 We have also observed the interstellar medium
around \3c at millimeter wavelengths with the MOPRA (Australia) and the
Swedish ESO Submillimeter (SEST) telescopes.
The  observation was targeted to probe the conditions of the ambient medium
by using the $^{12}$CO J=2--1 and $^{12}$CO J=1--0 line transitions.
In \S2, we summarize the observations.
In \S3, we describe our imaging results obtained with \chandra.
In \S4, we detail our spatially resolved spectroscopic study of the SNR.
In \S5, we present the results of the millimeter observations.
Finally, we discuss our results in \S6 and summarize our conclusions
in \S7. 

\section{Observations}
\3c\ was observed with the \chandra\ X-Ray Observatory for 66~ks
on 2001 September 6, with the back-illuminated chip S3 of
the Advanced CCD Imaging Spectrometer (ACIS-S,
G. Garmire\footnote{http://cxc.harvard.edu/proposer/POG/}),
at a focal plane temperature of -120$^o$. 
(For a review of the \chandra\ X-ray observatory, see Seward
2003\footnote{http://cxc.harvard.edu/cdo/about$_{-}$chandra/overview$_{-}$cxo.html}.)
The data were corrected for charge
transfer inefficiency (CTI) with tools provided by the ACIS instrument 
Team of Penn State University (Townsley \ea\ 2000).
The gain-map calibration was re-applied according to standard
CIAO 2.3 processing procedures.
Events with \asca\ grades (0, 2, 3, 4, 6)
were retained, and periods of high background rates were removed. The
total effective exposure time was 65.5~ks.
The spectral analysis was performed using $XSPEC~v11.2$\footnote{http://xspec.gsfc.nasa.gov}, and the spectra
were binned using a minimum of 20 counts per bin. 
When performing the spatially
resolved spectroscopic study (\S4), we subtracted a local background from
source-free regions within the S3 chip, and along the same galactic
latitude as the source region. This has the advantage of minimizing the
contamination by the galactic ridge emission. 
For the hot spot and the new point sources (\S3), we extracted
spectra from a ring surrounding the source.

The radio millimeter observations of \3c\
were performed in 1998 April  and 1999 March
% 9, 10, 11 and in April,
using the  Australian Millimeter Telescope
(MOPRA) and the Swedish ESO Submillimeter telescope (SEST) at La Silla
(Chile),
respectively.
The SEST observations  made use of the $^{12}$CO J=1--0 transition at 115.3 GHz and the $^{12}$CO J= 2--1 transition at 230.5 GHz.
To study the $^{13}$CO J=1--0 transition, we examined the International
Galactic Place Survey (IGPS)
data\footnote{http://www.ras.ucalgary.ca/IGPS/}.
The IGPS is an international consortium which, at cm wavelengths,
combines the Canadian Galactic Plane Survey (CGPS) using the Dominion
Radio Astrophysical Observatory's Synthesis Telescope in Canada,
the VLA Galactic Plane Survey (VGPS) using the National Radio
 Astronomy Observatory's Very Large Array in New Mexico, 
and the Southern Galactic Plane Survey (SGPS)  using the Australia
     Telescope National Facility's Compact Array in New South Wales.
The $^{13}$CO database was acquired with the
Five College Radio Astronomy Observatory in Massachusetts.
The images were processed using 
AIPS\footnote{Astronomical Information Processing Software,
\url{http://aips.nrao.edu/}}
 and the
Karma\footnote{\url{http://www.atnf.csiro.au/computing/software/karma/}}
 software (Gooch 1996). The results from the millimeter observations
are presented in \S5. 
 
\section{Imaging}

One of the main goals of the \chandra\ observation is to unveil the
nature of the hot spot.
To address this goal, we generated an energy color image.
We assigned a red color to the soft (0.5--1.5 keV) band,
a green color to the intermediate (1.5--2.5 keV) band,
and a blue color to the hard (2.5--10 keV) band.
This method has shown to be a powerful tool to resolve the hard non-thermal emission expected from a Crab-like pulsar or PWN from the  thermal X-ray emission
associated with the SNR.
In Fig.~1, we show the resulting image.
The central `hot spot' is clearly soft and does not correspond to a point source.
In Fig.~2, we show the hot spot on the same scale as a hard point source
discovered south-east of the SNR (\S3.1).
The morphology and spectrum of the hot spot
are similar to other knots seen in the SNR,
therefore we rule out its nature as a compact stellar remnant or a PWN
(\S4.3).

\subsection{New Point Sources}
We identify two new point sources in the \chandra\ field.
The first is 
a soft source located at the northeastern edge of the SNR
and appearing as a red source in Fig.~1 (see also Fig.~3).
The source is located at $\alpha$=19$^h$ 07$^m$ 38$^s$.3,
 $\delta$=+07$^o$ 09$^{\prime}$ 22$^{\prime\prime}$.9 (J2000)
 with a count rate of
%  Noticed channels     2 to    19
%  File observed count rate    6.1057E-03+/-3.18726E-04 cts/s( 97.1% total
%  Source file counts :    412.0
(6.1$\pm$0.3)$\times$10$^{-3}$ \cps in the 0.5--3.0 keV range.
Since its derived column density ($N_{\rm H}$$<$0.4$\times$10$^{22}$~cm$^{-2}$,
2$\sigma$) is much smaller than that towards \3c\ 
($N_{\rm H}$=3.2$^{+0.2}_{-0.3}$$\times$10$^{22}$~cm$^{-2}$ using the
\asca\ data, Safi-Harb et al. 2000),
 we rule out its association with the remnant.

Another interesting point source 
located just outside the south-eastern side of the SNR stands out as a hard (blue) source
in Fig.~1 (see also Fig.~3).
This source, located at $\alpha$=19$^h$ 07$^m$ 41.$^s$298;
 $\delta$=07$^o$ 06$^{\prime}$ 50.$^{\prime\prime}$98 (J2000) (with a
90\% error radius of 2.$^{\prime\prime}$2) 
 will be designated as CXO J190741.2+070650 hereafter.
In Fig.~2, we show a zoomed image of this new source shown on the same
scale as the hot spot.

It has been suggested earlier that this source is associated with \3c\ (Keohane 
et al. 2003).
In order to compare its spatial characteristics with 
\chandra's point-spread function (PSF), we
generated a PSF image at an offset angle of 2$^{\prime}$.2 (the location of the hard
source on the S3 chip) and at an energy of 5~keV
(characteristic of the source's energy histogram). We subsequently normalized the
PSF image to the source counts (1815 background subtracted source counts in the 0.3--8.0 keV
band), and used the PSF as a convolution kernel when fitting the source.
A two-dimensional gaussian model ($gauss2d$ in $SHERPA$ version 3.0.1) 
yielded a FWHM of 2.19
pixels, which is comparable to the PSF's FWHM of 2.03 pixels. Therefore, we conclude that
the hard source is consistent with a point source.
While a more detailed study of this source will be deferred to another paper
(in preparation), we show its spectrum  in \S4.4
and conclude that it is unlikely to be associated with \3c\ (see \S4.4).
%based on its larger column density, location, and unusual spectrum.

\subsection{SNR}

The \chandra\ broadband (0.5--10 keV) intensity image (Fig.~3) 
has an overall morphology similar to the \rosat\ (0.5--2.4 keV) image: 
a box-like shape with two lobes and enhanced X-ray emission from the
western lobe and the central `hot spot'. 
As well, two diffuse `jet'-like structures  appear east
and west of the hot spot running along the symmetry axis
of the SNR (in the southeast-northwest direction).
The \chandra\ image also reveals for the first time 
bright arcsecond-scale knots or clumps, most
notably in the western lobe and the northeast (Fig.~1).
As shown in Fig.~4, there is a close resemblance between the 
X-ray and the radio morphology,
except for the central X-ray `hot spot'.
In Fig.~5, we show the hard (4--6 keV) image with the radio contours overlayed.
This energy range was chosen to suppress the contribution from the
ejecta which give rise to emission lines outside this range (\S4).
While it was not possible to resolve this hard emission with \asca,
\chandra\ shows that it is contained within the radio
`shell'. We also note that the bright knotty structure of the SNR evident in
the soft band disappears in the 4--6 keV band.

\3c\ does not have the limb-brightened circular
morphology expected from a young SNR 
propagating in a homogeneous medium. It is rather centrally-filled with the
outer shock front missing or disrupted, particularly in the east;
or deformed, particularly in the west.
The western and south-western edge
is strikingly sharp and bright (in both the radio and X-ray images)
and looks almost rectangular in the south-west; indicating
strong shock conditions.
It then becomes less defined and softer 
towards the south-east and east, 
and reveals a soft a plume-like structure (regions 1 and 2, Fig.~6), 
indicating propagation in a less dense ambient medium.
Noticeably, there is a low-surface brightness `strip' running from north to south
in the western side of the remnant (between the `hot spot' and the western lobe;
regions 7, 8, 9; Fig.~6). This depression in X-ray emission also corresponds to faint
radio emission (Fig.~4).

\section{Spatially resolved spectroscopy}

In the following, we perform a spatially resolved spectroscopic study on the remnant
guided by our imaging study.
This study is targeted to address the morphological differences between
the eastern and western sides of the remnant and to resolve the soft
and hard X-ray emission from the remnant.
As we found with \asca, 
the \chandra\ spectrum of \3c\ is dominated by thermal emission with 
emission lines from Ne and the Fe-L blend, Mg, Si, S, Ar, Ca, and Fe.
To address the morphological differences between the eastern and western lobes
and characterize the spectrum of the central `hot spot',
we have first extracted spectra from the eastern and western lobes and
the hot spot (see Fig.~6). 
Furthermore, to resolve the soft and hard components of the global \asca\ fit,
we extracted spectra from 12 arcseconds-scale regions, as shown in Fig.~6.
(The colors of the regions' boundaries are selected to remind the 
reader of the energy color image, Fig.~1).

\subsection{One-component models}
We first fitted the spectra with one-component thermal models. These include
collisional ionization equilibrium models such as $mekal$
(Mewe, Gronenschild, \& van den Oord 1985; Liedahl et al. 1990),
and non-equilibrium ionization (NEI) models which are more appropriate
for modeling young SNRs in which the plasma has not yet reached ionization
equilibrium.  We used the following NEI models (Borkowski, Reynolds,
\& Lyerly 2001) in $XSPEC~v11.2$:
\begin{itemize}
\item{
\vshock, a NEI model (with variable abundances) which
comprises a superposition of components of different 
ionization ages appropriate for a plane-parallel shock. 
This model is characterized by the constant electron temperature
($T$) and by $\tau$, the ionization timescale which characterizes
the shock ionization age, and which is
given by $\tau$=$n_{e}t$; where $n_e$ is the post-shock
electron density and $t$ is the time since the passage of 
the shock;}
%
%$T$, and the shock ionization age, 
%$n_{0}t$ (where $n_0$ is the pre-shock density, 
%and $t$ is the age of the shock). In this paper, we }
\item{$vsedov$, a  NEI model (with variable abundances)
which follows the time-dependent ionization of the plasma 
in a SNR evolving according to the Sedov self-similar dynamics.
Since this model includes a range of temperatures and ionization
timescales, it is more appropriate for fitting the large regions
(eastern and western lobes).}
\end{itemize}
We found that none of the single-component models give adequate fits to the spectra
as they do not account for the emission above $\sim$4~keV.
For example fitting the hot spot with a one-component \vshock\ model
with variable metal abundances 
yields a  reduced $\chi^2$ value of $\chi^2_{\nu}$=1.25 ($\nu$=118
degrees of freedom) with all the residuals found above $\sim$3~keV.
For the eastern and western lobes, we obtain $\chi^2_{\nu}$
$\geq$10 even with the $vsedov$ fit which includes
a range of temperatures and ionization timescales.

We also find that single-component models fail to account for 
the emission above 4~keV for all the 12 selected regions despite
their small sub-arcminute size;
and that at least another hot component should be added.
This indicates that the spectrum of \3c\ is 
more complex than in any other known SNR, 
and suggests that there is mixing of shocked ejecta and circumstellar material
on small spatial scales.

\subsection{Two-component models}

We subsequently fitted the spectra of the eastern and western lobes
and the hot spot with a two-component thermal model.
 This was motivated by the failure of one-component models (\S4.1) and the
expectation of a high- and a low-temperature plasma 
associated with the supernova blast wave 
and reverse-shocked ejecta.
In Table~1, we summarize our fits using a simple two-component thermal bremsstrahlung
(TB) model. Gaussian lines were added to account for the emission
from Ne, Mg, Si, S, Ar, Ca, and Fe.
This model allows us to identify the centroids of the  line emission and determine the
 column density and the
parameters of the continuum emission, $independently$ 
of the abundances.
As shown in Table~1, 
the centroids of the strong Fe-K line detected in the eastern and western lobes 
are lower than the values expected
from plasma in collisional ionization equilibrium at the fitted
temperatures. This indicates that the hot plasma has not
yet reached ionization equilibrium and should be 
instead characterized by a NEI model --
a result that is
consistent with the global \asca\ fit (Safi-Harb et al. 2000).  
Furthermore, when comparing the eastern to the western lobe,
we find that the column density is lower, the fitted temperature of
the hot component  is higher,
and the centroid of the Fe-K line  is at a lower energy
indicating a smaller ionization timescale for the eastern lobe.
We interpret this as evidence for a lower overall density in the eastern side of the
remnant.

The above findings from our TB fits are consistent with our NEI model fits 
discussed below.
In Tables~2 and 3, we summarize the two-component \vshock\  fit to the hot spot, 
the eastern and western lobes;
and in Figures 10 and 11, we show their spectra fitted with these models. 
Examining the \vshock\ parameters,
we note that while the absolute values of $N_{\rm H}$ and $kT$ are not exactly the same
as for the two-component TB model (Table~1), 
their relative values are similar, namely:
$N_{\rm H}$, $kT_s$, and $\tau_h$ are smaller in the eastern lobe, and
$kT_h$ is higher for the eastern lobe.
We caution the reader
that the fits to the eastern and western are not satisfactory
($\chi^2_{\nu}$$\geq$2; see the residuals in Fig.~11).
This is expected in  a multi-temperature, multi-ionization timescale plasma.
We show below that the spectral parameters vary within these large regions;
so the parameters shown in Table~3 should not be taken at face value,
but they should be adequate in describing 
the global spectral properties in the eastern and western
halves of the remnant.
For the hot spot, the soft component
is equally well described by a \vshock\ model with a high ionization timescale
or a $mekal$ model.
This indicates that the soft component
is approaching (or has already reached) ionization equilibrium.

In Table~4, we present our two-component \vshock\ fits to the
12 selected small scale regions shown in Fig.~6; and in Figs 12--15,
we show the spectra of some regions of interest fitted with the best
fit \vshock\ models. (We chose to show representative regions from the
eastern and western lobes, the interior, the low-surface brightness
strip, and the eastern clumps.)
First we note that average temperatures and ionization timescales are consistent
with our global \asca\ fit to the SNR. 
However, the excellent spatial resolution of \chandra\ revealed
some spectral variations on small scales. 
A clear outcome from our spatially resolved spectroscopic study
is a general trend of the column density $N_{\rm H}$ increasing
from $\leq$3.0$\times$10$^{22}$~cm$^{-2}$ in the east (regions 1 and 2)
to $\geq$4.0$\times$10$^{22}$~cm$^{-2}$ in the west
(region 12, see Fig.~7).
For the soft component,  the temperature varies from $\sim$0.15 keV on the eastern side (regions 1 and 2) to $\sim$0.26 keV in the low-surface brightness
 `strip' in the west (regions 7 and 8, see also Fig.~8).
The ionization timescales are of the order of $\sim$few$\times$10$^{11}$--
5$\times$10$^{13}$~cm$^{-3}$~s (except in regions 7-9
in the low-surface brightness `strip'), 
indicating that the soft plasma is close to ionization equilibrium
in most regions.
For the hard component, the temperature is $\sim$1--3~keV
and is highest (although poorly
constrained) along the eastern jet-like structure (region 3, see also Fig.~9).
(The high temperature derived for the latter region could be
indicative of the presence of a third non-thermal component).
Furthermore, except for the interior emission located in the low-surface
brightness `strip', the ionization timescales of the hard component
are lower than those of
the soft component, and of the order of $\sim$~5$\times$10$^{10}$--10$^{12}$~cm$^{-3}$~s,
indicating that the hot plasma has not yet reached ionization equilibrium
in most regions.
 
\subsubsection{Abundances}
We now discuss the abundances derived from our fits to determine the ejecta
and blast wave temperatures and address the two scenarios proposed
in our previous work (see \S1).
First, when we freeze the abundances of both \vshock\ components to their solar values,
we find that these models yield unacceptable fits,
confirming the presence of ejecta.
To determine whether the hard component arises from the blast wave,
we freeze the hard component's metal abundances to their solar values and allow
the abundances of the soft component to vary as needed.
While the fits improve dramatically especially near
the Ne, Mg and Si lines, they do not account for the emission from S
and particularly from Fe near 6.5 keV; and require
enhanced S and Fe abundances.
This suggests that the hard component is (at least partially) due to shock-heated
S and Fe ejecta. 

The apparent high S abundance might be an artifact of the models
used because the S line falls in the energy range where the X-ray
spectra from the low- and high-temperature components overlap.
As for the high Fe abundance, we note that the hot
temperatures derived for the eastern and western
lobes from the two-component \vshock\ models 
(Table 3)
are less than those derived from the thermal bremsstrahlung
fits (Table~1) or the fits to the hard component only (see below). 
One might then argue that the high Fe abundance could be also
an artifact of the fits, since underestimating the
hard component's temperature  would then underestimate the
continuum emission and subsequently overestimate 
the line emission from Fe. To test for this and to determine
the abundances of the hard component independently of the
soft component, we fitted the 4-8 keV band with the \vshock\ model.
We find that for both the eastern and western lobes,
an above-solar Fe abundance is needed.
For example, the 4-8 keV spectrum of the eastern  lobe
yields $kT_h$ = 5.1 (3.2--16) keV,
an ionization timescale $\tau_h$ = 2.95 (2.5--5) $\times$ 10$^{10}$~cm$^{-3}$~s,
and an abundance of Fe$_h$  = 11.8 ($\geq$6)~\sol\ (errors are 2$\sigma$).
For the western lobe, we find $kT_h$ = 2.8 (2.2--3.7) keV,
$\tau_h$ = 5.2 (4.5--6.3) $\times$ 10$^{10}$~cm$^{-3}$~s,
and Fe$_h$ = 9.4 (6.8--13.8)~\sol.
Therefore, regardless of the model used, a high Fe abundance is needed 
to account for the hard X-ray emission, confirming an ejecta origin.

Now to determine the origin of the soft component, 
we subsequently froze the soft component abundances to solar and allowed
the abundances of the hard component  to vary.
We find that regardless of the hard component metal abundances,
the spectra of many regions do not fit well in the 0.5--4 keV range.
For example, the global fits to the eastern and western lobes require
enhanced O, Ne, and Ca abundances. Furthermore, the bright northeastern
clump (region 10) requires enhanced Ne and Ca, and the southwestern
knots require enhanced O, Ne, and Ca.
This also suggests that the soft component is partially due to shock-heated
ejecta. 
As shown in Table~3, the best fits for the eastern and western lobes
are achieved when tying the abundances of the
soft and hard component, except for Fe.

For the hot spot,
the hard component requires a S abundance of
$\geq$3~\sol\ (see Table~2). Since its spectrum is soft, there aren't
enough counts at higher energies and no emission was seen from Fe-K. 
The abundances of the soft component are however consistent with
solar values.
The enhanced brightness, the low temperature,
the large ionization timescale and the solar abundances
of the soft component all
indicate that its soft X-ray emission arises from a shocked cloudlet (\S6).
The high S abundance suggests an ejecta origin; however, as mentioned
above,
this could be an artifact of
the two-component \vshock\ fit.

In Table~4, we summarize the S and Fe abundances
determined from our two-component \vshock\ fits to the smaller
scale regions. 
An above-solar S and Fe abundance is needed in most regions 
(except in the interior regions: the western `jet', region 4, and the
northwestern faint, region 7).
We caution the reader that the parameters of these fits are poorly constrained
due to the large number of parameters and the poor signal to noise
ratio especially in the hard energy band.
The errors shown in Table~4 are 2$\sigma$ and have been determined after freezing
the other component's parameters to their best values.

\subsection{Multi-component model}
It is highly likely that a two-component NEI model is too simple or inadequate
to describe this SNR and that a more complex multi-component model is
needed. This argument is supported by the residuals seen
in the spectra, and  is not unrealistic for a SNR which results from
the explosion of a massive star in a complex environment 
and near a molecular cloud. 
Simulations of off-centered supernova explosions in pre-existing
wind bubbles have shown that, in a dense medium, the
 SNR could become elongated
parallel to the progenitor's direction of motion and that jet-like features
could develop along the same direction (Rozyczka et al. 1993). 
The elongated morphology of \3c, the jet-like structures seen along
the symmetry axis, and the evidence of a shell-like structure
around \3c\ (\S5.2) support this model. 
Such a complex environment would then give rise
to a  multi-component plasma arising from reverse-shocked ejecta, 
shocked ambient and circumstellar medium, 
and from the reflected shock(s) from the molecular cloud.

An additional non-thermal component could also arise from strong
shocks resulting from the interaction of the blast wave or reverse shock
with a dense medium, or from highly relativistic outflows emanating e.g.
from a hidden compact object.
Hints of non-thermal emission from \3c\ exist from the
high-temperatures ($kT_h\geq$2~keV) derived for some regions, in
particular for the interior eastern `jet' (region 4, see Table~4).
However, the poor signal to noise ratio in the hard band,
the presence of the Fe-K lines near 6.5~keV indicating a 
(partially) thermal origin,
and the very large number of free parameters
in the multi-component model prohibit us from
a further investigation. A more detailed study has to await deeper
\chandra, \xmm, and \astroe\ observations.
We here conclude that the plasma in \3c\ is dominated by a
multi-temperature,
multi-ionization timescale shock-heated ejecta mixed with
shocked ambient/circumstellar material.

To put an upper limit on the luminosity of a 
hard non-thermal component that could
be present in \3c, we determine the integrated
flux from the SNR regions in the 4--6 keV energy band.
As we mentioned before (\S3.2), this energy band should be free
from line emission from the SN ejecta, so an underlying
hard non-thermal component would contribute to the
continuum emission.
Our fits to the hard band only yield an integrated flux
of $\leq$ 1.1$\times$10$^{-13}$~erg~cm$^{-2}$~s$^{-1}$,
which at a distance of 10~kpc translates to
$L_x$ (4--6 keV) $\leq$ 1.3$\times$10$^{33}$ erg~s$^{-1}$.
This upper limit is consistent with the limit determined
in the 5--15 keV range
using the combined \asca\ and \rxte\ spectrum (Safi-Harb et al. 2000).

\subsection{The hard point source, \cxo}
When extracting the spectrum of \cxo, we chose the background from a ring
surrounding the point source.
The total background-subtracted count rate from \cxo\ 
is 2.88$\times$10$^{-2}$ \cps\ in the 1.0--10.0 keV
energy range, yielding a total of 1,922 counts.
 As shown in Fig.~16, the spectrum of the source is heavily absorbed, 
hard, and has a prominent Fe-line near 6.4 keV. 
Fitting the spectrum with an absorbed power law model plus a narrow Gaussian
 line yields an adequate fit with the parameters summarized in Table~5.
 The corresponding unabsorbed flux is 2.1$\times$10$^{-12}$~erg~cm$^{-2}$~s$^{-1}$
in the 1--10 keV energy range. 

A more detailed study of this source is beyond the scope of this paper
and will be presented elsewhere.
%We note here  that the source
%is unlike any known rotation-powered,
%accretion-powered, magnetically
%powered neutron star, or a cataclysmic
%variable. 
We note here that it is highly unlikely to be
associated with \3c\ due to the following reasons:
1) its location outside \3c\ (a velocity of $\geq$1,300~km~s$^{-1}$ 
is required for the source to be associated with
the $\sim$5,000 year-old SNR at a distance of 10 kpc, see \S6);
2) its high interstellar column density
($\sim$0.6$\times$10$^{22}$~cm$^{-2}$ higher than that of region 11, Fig.~6);
and
3) its X-ray spectrum which is unlike typical rotation-powered,
accretion-powered, magnetically powered neutron stars, or  cataclysmic
variables.
Its spectrum is however
not atypical of Seyfert II Active Galactic Nuclei (AGN) which display 
heavily absorbed spectra with strong Fe lines. 
Assuming the Fe line is a red-shifted fluorescence line,
the derived distance is 40$\pm$20~Mpc, and the corresponding luminosity is 
$\sim$4$\times$10$^{41}$~erg~s$^{-1}$ (at a distance of 40 Mpc).

\section{Millimeter Observations}
\subsection{The $^{12}$CO observation}
We first observed \3c\ with MOPRA using six different locations 
along a line running from east to west and with
a short integration time on each point of 3~minutes. We found
an enhancement in emission on the western side at a 
velocity $v\sim$40~km~s$^{-1}$.
We subsequently
pointed the SEST telescope towards the center of the SNR
and made a 9$\times$9~pixel map, with 45$^{\prime\prime}$ pixel to 
pixel separation and a central velocity of 40~km~s$^{-1}$.
The antenna beam widths were 45$^{\prime\prime}$ and 22$^{\prime\prime}$ for the 
$^{12}$CO J=2--1 and J=1--0 transitions,
respectively. The observations were carried out using a position beam switching mode,
choosing an absolute reference position of $\alpha$ = 19$^h$ 14$^m$ 18$^s$.3 and
$\delta$ = +07$^o$ 46$^{\prime}$ 16$^{\prime\prime}$, J2000.
Such a distant location was required to find a region without radio emission in this crowded field.
The receivers yielded an overall system temperatures (including sky noise)
of $\sim$340K at 230 GHz and $\sim$400K at 115 GHz, while the receiver temperatures were around 100K at 230 GHz
and 10K at 115 GHz. The back end was an acousto-optical spectrometer. 
The low resolution spectrometer with a channel separation of 0.7 MHz 
(or 1.8 km~s$^{-1}$ and 0.9  km~s$^{-1}$ velocity resolution
for  the 1-0 and 2-1 transitions, respectively)
was used for these observations.
All the spectra intensities (antenna temperatures) were converted to the
main brightness  temperature scale, correcting for a main beam 
efficiency of 0.7 at 115 GHz and 0.5 at 230 GHz.
The pointing accuracy, obtained from measurements of the SiO maser W Hya and IRAS 15194,
 was better than 5$^{\prime\prime}$.

As shown in Fig.~17 and in Table~6,
there is an enhancement in the
molecular gas emission towards the west that is correlated with
the enhanced X-ray emission. While the millimeter line emission
does not seem to have the
morphology of a solid flat `wall', as would be expected to fit the shape
of the western side of \3c, this gas distribution is
consistent with the eastern half of the SNR being more tenuous in X-rays,
since it appears to be expanding in less dense gas. A correspondence can
be observed between the maximum in the $^{12}$CO J=1--0 emission, centered near $\alpha$ (J2000)= 19$^h$ 07$^m$ 34$^s$, 
$\delta$ (J2000) = 07$^o$ 10$^{\prime}$ 15$^{\prime\prime}$, 
and a small indentation in the X-rays.
In Fig.~17 (right), we show the comparison of the X-ray emission with the 
$^{12}$CO J=2--1 distribution. 
It shows essentially the same morphology as the $^{12}$CO J=1--0
transition, with
a smooth gradient from west to east.
Based on these correlations, we assume that the SNR \3c\  is associated with a molecular cloud with a mean velocity of $\sim$40 km~s$^{-1}$; 
and using the Galactic rotation curve, 
we derived two possible distances for \3c:
2.3 $\pm$0.1 kpc and 10.3$\pm$0.2 kpc for the close and far distances, respectively.
The latter distance is consistent with previous distance
estimates to \3c\ (see \S1).

\subsection{The $^{13}$CO J=1--0 line transition}
To confirm the association between \3c\ and the molecular cloud found
above, we examined the $^{13}$CO J=1--0 line transition using the
IGPS data.
Figure~18 displays the distribution of the $^{13}$CO J=1--0 line in the environs
of \3c,
for several velocity channels around $v\sim$40 km~s$^{-1}$. The gas distribution
is displayed in greyscale and 
the contours represent the \chandra\ X-ray image of \3c\ 
(smoothed with a 3$^{\prime\prime}$ gaussian). 
%In this representation West is up and North is to the left.

A filamentary molecular feature is noticed nearby 3C\,397 between $v\sim$35.4
km~s$^{-1}$  and $v\sim$~41~km~s$^{-1}$. 
Between $v\sim$35.4 and 36.3~km~s$^{-1}$, approximately, the
molecular cloud partially overlaps the SNR. 
From $v\geq$36.3 to $\sim$38.4 km~s$^{-1}$,
dense gas surrounds the western and southern 
(up and right sides) borders of the SNR, while the eastern extreme 
(bottom side in Fig.~18) expands into lower density  medium, 
in concordance with the results shown in the $^{12}$CO lines
(\S5.1).
Around $v\sim$39 km~s$^{-1}$ and up to $v\sim$41.8~km~s$^{-1}$,
traces of an open shell around 
\3c\ are apparent. In Fig.~19, we show the image in the $^{13}$CO J=1--0
line transition with the velocity integrated from 35.4 to
41.3~km~s$^{-1}$. The apparent shell-like structure around \3c\ could
be a relic of the wind-blown bubble created by the precursor star.
Higher resolution molecular and atomic observations are highly desirable to
fully explore the interaction of this SNR with its surrounding
interstellar gas.  

\section{Discussion}

\subsection{Distance}
We now estimate the distance to the SNR using our fitted column density to the remnant.
We use the range $N_{\rm H}$ = (2.9--4.3)$\times$10$^{22}$~cm$^{-2}$ which encompasses
the values derived for all the regions (Table~4).
The extinction per unit distance in the direction of the system can be estimated
from the contour diagrams given by Lucke (1978): 
$E_{\rm B-V}/D$ $\sim$ 1.0 mag~ kpc$^{-1}$.
Using the relation $<N_{\rm H}/E_{B-V}>$ = 5.55 $\times$ 10$^{21}$~cm$^{-2}$~mag$^{-1}$
(Predehl \& Schmitt 1995), we derive a distance of 5.2--7.8~kpc.
This distance is smaller that that derived from the millimeter observations
(10.1--10.5~kpc),
but not inconsistent 
with the range (6.4--12.8 kpc) determined from HI absorption studies.
We will subsequently scale our calculations to the
parameter $D_{10}=D/10$ kpc, and keep
in mind that $D_{10}$ could span the range 0.6--1.3.

\subsection{A compact stellar remnant and the nature of the hot spot}
We found two point sources in the \chandra\ field: a soft source
at the north-eastern edge, and a hard source just outside the
south-eastern edge. As discussed in \S3 and \S4.4,
we believe that none of these sources is associated with the SNR.

The hot spot was proposed to be the compact stellar remnant or a mini-plerion.
Our imaging and spectral analysis has confirmed that the hot spot can not be either.
 We can not however rule out
the presence of a compact object buried underneath the thermal emission
with an unabsorbed flux of
$\leq$~6$\times$10$^{-13}$~erg~cm$^{-2}$~s$^{-1}$, which translates to
a luminosity of $L_x$ (0.5--10 keV) = 7$\times$10$^{33}$~$D_{10}^2$
erg~s$^{-1}$.

As to the origin of the hot spot, we suggested in \S4.2.1 that it
could be a clump arising from a shocked cloudlet.
To estimate the corresponding density, we
use the \vshock\ model parameters summarized in Table~2.
The observed measured emission measure of the soft component ($EM_s$) 
of 0.108 corresponds to
 $\int n_e n_{\rm H} dV$ $\sim$ $f_s n_e n_{\rm H} V = 10^{14} (4 \pi D^2) (EM)$ = 
1.3$\times$10$^{59}$~cm$^{-3}$; where $f_s$ is the volume
filling factor of the soft component, $n_e$ is the post-shock
electron density, and $n_{\rm H}$ $\sim$ $n_e$/1.2.
Using a radius of 12$^{\prime\prime}$.5 for the hot spot, we estimate
$n_e$~=~72~$f_s^{-1/2}$~cm$^{-3}$, implying an 
upstream ambient density $n_0$ = $n_e/4.8$ = 15~$f_s^{-1/2}$cm$^{-3}$
(for cosmic abundance plasma and the strong shock Rankine-Hugoniot jump
conditions; $n_0$ here includes only hydrogen).
This corresponds to $n_0$$\sim$15--50~cm$^{-3}$ for 
filling factors $f_s$$\sim$0.1--1.0.

\subsection{Origin of the soft and hard components}

In the following, we discuss the origin of the soft and hard components;
using our two-component model fits (\S4.2).
We again remind the reader that a two-temperature component model is likely
to be an oversimplified model for this complex SNR (\S4.3), so
considerable caution is required in interpreting the two-component
model fits. The parameters derived from our two-component fits
nevertheless describe the global spectral differences across the
SNR.

In our \asca\ study we suggested two scenarios for the interpretation of the
two-component global fit. 
For the 2~kyr-old ejecta-dominated SNR, the soft component would arise from 
reverse-shocked ejecta and the hard component from the blast wave shocking an ambient
medium with a density of $n_0$$\sim$~1.0~$D_{10}^{-1/2}$~cm$^{-3}$. 
For the 5~kyr-old SNR propagating in a dense medium, the soft component
would correspond the blast wave propagating in a medium of 
density $n_0$$\sim$33~$D_{10}^{-1/2}$~cm$^{-3}$.
In the first scenario, one expects the hot component to  
originate from the SNR `shell'
and to be characterized by solar (or sub-solar abundances);
while the soft component should be located  inside the SNR `shell'
and characterized by above-solar abundances.
For the second scenario, one expects the opposite behavior.
Furthermore, the ionization timescales should decrease 
with radius for the blast wave component (forward shock) since the interior
is shocked first, and to increase with radius for the ejecta
component (reverse-shock).

Morphologically, the \chandra\ energy color image (Fig.~1) 
shows that the softest
X-ray emission is found near the edge of the eastern lobe and the south-eastern
boundary, and that the hard component (while
visible in most regions) is more concentrated inside the SNR (see also Figs
5, 8, 9). Furthermore, our
spectroscopic study shows that 
the ionization timescales have a general trend to decrease outwards for the
soft component and to increase outwards for the hard component 
(except for the regions inside the low-surface brightness `strip',
see Figs 8 and 9).
This, together with the high Fe abundances required to fit the hard
component, support the second scenario 
where the SNR is propagating in a dense medium.
Using our \vshock\ fit to the soft component (Tables 2-4) and following
the calculations done above for the hot spot (\S6.2), we estimate
pre-shocked ambient densities
ranging from  $n_0$$\sim$15--70~$(f_sD_{10})^{-1/2}$~cm$^{-3}$, a range that
brackets the average density derived from our global \asca\
fit for large filling factors,
$f_s$.

For the hard component, the derived high Fe abundances, 
relatively low ionization timescales and emission measures support
the presence of reverse-shocked Fe bubbles. Using
our \vshock\ fits to the hard component (Tables 2-4), we 
estimate pre-shocked electron densities, $n_e$, 
ranging from 1.5--8~$(f_hD_{10})^{-1/2}$~cm$^{-3}$
which then imply that the hot plasma has been shocked 
at a time $t$=$\tau_h$/$n_e$$\sim$700--4,000 $(f_hD_{10})^{1/2}$~years
ago. The volume filling factor, $f_h$, can not be accurately determined
due to the poor signal to noise ratio in the
hard energy band.
For a filling factor $f_h$$\sim$0.1--1.0, we estimate
$n_e$$\sim$1.5--25~cm$^{-3}$ and $t$$\sim$220--4,000~years;
a reasonable timescale for a reverse shock to develop in a 
$\sim$5,300 year-old SNR.
Regions 7, 8, and 9 are an exception: 
the soft and hard component's ionization timescales are different
from the other regions. 
This, together with their low-surface radio and X-ray brightness,
suggest that the conditions in this region are different
from elsewhere in the SNR.
Furthermore, the much lower ionization timescales of the soft component (Table~4) 
indicate that the cool plasma has been just recently shocked in these regions. 
We speculate that this is due to a reflected shock,
perhaps from encountering a nearby molecular cloud.
Since the spectral parameters are poorly constrained,
a better understanding of the ambient conditions
and a theoretical investigation of the SNR dynamics
has to await future X-ray observations.

\subsection{Classification of \3c: Interaction with a cloud?}
As we mentioned above, our $^{12}$CO and $^{13}$CO study shows
that the ambient medium is more
tenuous in the eastern lobe, and indicates the
presence of a molecular cloud near $\sim$40~km~s$^{-1}$.
The peculiar rectangular morphology, the sharp western boundary, the
enhanced radio, millimeter, and X-ray emission and the
higher column density in the west support the scenario
where \3c\ is expanding in an inhomogeneous medium, being denser
in the west; and 
likely encountering or interacting with the molecular cloud.
While the absolute value of $N_{\rm H}$ is model dependent, the
 difference in $N_{\rm H}$ between the east and west is not (Tables~1-4).
 If the difference of $\sim$(0.5--1.0)$\times$10$^{22}$~cm$^{-2}$ 
 is solely due to a density gradient caused by a molecular cloud, then
the corresponding average density would be $\sim$ 160--320~
($\delta$$l_{10~pc}$)$^{-1}$~$D_{10}^{-1}$~cm$^{-3}$;
where we have assumed that the line of depth
of the SNR in the cloud, $\delta$$l$, is close to the average
SNR size of 10~pc.
% (7--13 pc).

\3c\ bears resemblance to mixed-morphology SNRs like 3C\,391 (Chen \& Slane 2001)
and CTB~109 (Rho \& Petre 1997)
 because of the following:
1) it has a centrally filled X-ray morphology dominated by thermal emission;
2) it has a `shell'-like radio morphology, and 3) it 
displays asymmetries in the overall X-ray morphology and
variations in
the distribution of $N_{\rm H}$ across the remnant which have been attributed
to the presence of dense material that the remnant is encountering.
In mixed-morphology SNRs,  the derived abundances from X-ray spectra
are however more or less consistent with solar values,
indicating that their enhanced X-ray emission is due
to either evaporation of engulfed cloudlets or a radiative stage of evolution 
in which thermal conduction plays an important role.

For \3c\ however, 
the X-ray spectrum is more complex because of: a) the need for a two-component
model everywhere in the remnant and in arcseconds-scale regions,
b) the enhanced metal abundances
 indicating an ejecta origin, 
and c) the presence of an outer X-ray `shell'
which is strongly correlated with the radio `shell'. 
We note here however that this outer edge has a geometry
that is more rectangular than spherical
and that some parts of the remnant's limb are missing.
Furthermore, even the outermost western boundary (region 12) hints at the
presence of shocked ejecta Fe 
indicating that the reverse shock did penetrate to
large radii (Table~4). This could occur in the presence of
vigorous turbulence caused by the presence of Fe bubbles.
Three-dimensional hydrodynamical simulations by Blondin et al. (2001) 
show that in this case, the Fe bubbles will be mixed with
other heavy elements and with the ambient normal abundance gas.
Furthermore, turbulence will affect the morphology of the SNR 
by creating `jet'-like features resulting from the interaction of
bubbly ejecta with the ambient medium, and by enhancing the
radio synchrotron emissivity in strong turbulent regions.

While the above properties indicate that \3c\ is unlike the 
`classical' mixed-morphology
SNRs, the differences can be explained if we take into account 
the relative youth of \3c\  compared to the other
SNRs in which the signature of ejecta has been wiped out due to a later
evolutionary stage in the SNR lifetime.
An SNR enters the radiative pressure-driven snowplow phase (PDS) when its radius is:
$r_{PDS}$ = 14~$E_{51}^{2/7}$~$n_0^{-3/7}$ (pc).
Using an explosion energy $E_{51}$$\sim$1.2~$D_{10}^{5/2}$ and an average ambient density $n_0$$\sim$33~$D_{10}^{-1/2}$ (Safi-Harb et al. 2000),
we estimate a radius $r_{PDS}$$\sim$3.3~pc which is comparable
to the size of the remnant ($R$$\sim$3.5-7~pc).
This implies that \3c\ must be entering into its PDS phase; and so
we speculate that it will eventually resemble the mixed-morphology SNRs.

It is currently believed that the excitation of the
hydroxyl radical 1720 MHz maser is
a good indicator of interaction between SNRs with dense molecular
environments.
Evidence that mixed-morphology SNRs produce OH masers
has been recently accumulating (e.g. Yusef-Zadeh et al. 2003).
To search for maser emission from \3c, we have observed it
at 1720 MHz with the VLA on 2003 May 16 (W. M. Goss, private communication).
The beam was 50$^{\prime\prime}$ by 35$^{\prime\prime}$,  the 
rms noise about 7 mJy/beam (comparable to
the sensitivity achieved in other searches; W. M. Goss, private communication)  and 
the velocity resolution of  about 1.2 km/s over 127 channels.
No lines were found. However, this does not necessarily imply the absence
of interaction with a molecular cloud, but could rather 
indicate that the critical conditions required to excite OH masers are not yet achieved (Green et al. 1997).
Similarly, we find that the  $^{12}$CO 2-1/1-0 ratio (after converting
the antenna temperature into brightness temperature) is
quite homogeneous throughout the field, and about 0.87. 
Thus, this is not a very strong indication that some specific cloud is being shocked,
as the expected ratio for shocked gas is $\geq$1.
In conclusion, the lack of OH masers and the small ratio of 
CO~2--1/1--0 ($<$ 1) could indicate that the molecular cloud 
in the close vicinity has not yet been overrun by the shock wave. 
Alternatively, if the shock/cloud interaction is occurring in \3c, 
it must be taking place along the line of sight. 
This latter model provides an explanation to the X-ray shocked material
and the lack of OH (1720 MHz)
masers, because these masers are mainly excited when the shock front is
almost transverse to the line of sight 
(Wardle 1999).

A high-resolution
study of the molecular and atomic gas in the environs
are needed to better probe the level of excitation of the 
adjoining molecular cloud.

\section{Conclusions}

We have presented a \chandra\ ACIS-S3 observation of the SNR 3C\,397.
Previous studies of this SNR have shown that
the remnant has a double-lobed elongated
morphology with enhanced X-ray emission from
the center and the western lobe.
Most intriguing is a central enhancement of X-ray emission seen with 
\rosat\ and suggestive of
a compact X-ray source or a PWN associated with 3C\,397.
The spectrum of this `hot spot' could not be resolved with previous X-ray missions and
left the classification of 3C\,397 uncertain.
\chandra's imaging capabilities allowed us to
resolve the emission from the hot spot. We 
confirm its thermal nature, thus ruling out a PWN or
a central compact object.
We put an upper limit on the luminosity of a buried compact object 
 $L_x$ (0.5--10 keV) $\leq$ 7$\times$10$^{33}$~erg~s$^{-1}$ 
at an assumed distance of 10~kpc.
In addition, we discovered a hard point-source, CXO J190741.2+070650, 
located outside and southeast of the SNR. The properties
of this source lead us to believe
that it is not associated with 3C\,397,
and that it is most likely a nearby AGN.

The \chandra\ images reveal the softest X-ray emission from 
a plume-like structure at the edge of the eastern lobe and in the 
southeast, arcseconds-scale clumps and knots most notably in the
western lobe, jet-like structures
running along the symmetry axis of the SNR, a low-surface brightness
diffuse emission between the hot spot and the western lobe,
 and a sharp western
boundary which is almost rectangular in the southwest.
Except for the central `hot spot',
the clumps and knots seen in the
broadband X-ray image are strongly correlated with the radio VLA image.
The line-free 4--6 keV X-ray emission appears 
however to be contained within the radio shell.
 
Our spatially resolved spectroscopic study of the small scale regions 
shows that one-component
models are inadequate, and that at least two NEI thermal components
are needed to fit the spectra of each region.
In particular, we find that:
\begin{itemize}
\item{For the hot spot, the soft component has reached
ionization equilibrium and is
characterized by $kT_s$$\sim$0.18$\pm$0.3~keV and solar abundances.
The hot component is far from ionization equilibrium
and is characterized by $kT_h$$\sim$2.5$^{+1.0}_{-1.4}$~keV
and an enhanced S abundance ($\geq$3.0 at the 2$\sigma$ level).
We interpret the hot spot as a shocked cloudlet with a pre-shocked
density $\sim$15--50~cm$^{-3}$ and possibly mixed with hot S ejecta.}
\item{For the eastern and western lobes, our two-component
\vshock\ fits require enhanced abundances in both the
soft and hard components; in particular a high Fe abundance
($\geq$9~\sol) is needed to account for the strong Fe-K lines at
6.5~keV.
This indicates that the hard component is at least
partially due to shock-heated Fe ejecta.
The column density and the brightness are higher in the west
indicating a denser medium on the western side.} 
\item{For the small scale regions, we find that two-component
\vshock\ models yield adequate fits, with  a general trend
of increase of $N_{\rm H}$ from east to west.
The soft component's
temperature varies from $\sim$0.15 keV in the eastern and southeastern
edge to $\sim$0.26~keV in the low-surface brightness interior.
The hard component's parameters are less constrained but
the temperature is $\sim$1--3 keV and possibly higher
along the eastern `jet'-like structure. 
The ionization timescales of the soft component are generally
higher than those of the hard component. Furthermore,
the abundances of the soft component are generally consistent
with solar (except for some regions)
while those of the hard component generally require
above-solar S and Fe abundances. 
We interpret this as evidence for a combination of
a multi-temperature ejecta and shocked ambient/cicumstellar material.
The enhanced Fe abundances in the outermost regions of the
remnant indicate that the ejecta have penetrated to large radii
as would be expected from enhanced turbulence caused by shocked
Fe bubbles.}
\end{itemize}

The quality of the data does not allow us to constrain the parameters
of the hard component.
While we do not find compelling evidence for hard non-thermal emission 
from \3c, we put an upper limit on the luminosity
of the hard line-free emission in the 4--6 keV of
 $L_x$ (4--6 keV)$\leq$1.3$\times$10$^{33}$~$D_{10}^2$
erg~s$^{-1}$.

The overall imaging  and spectral properties
of \3c\ favor the interpretation of the SNR being a $\sim$5,300 year-old
SNR propagating in a dense medium, and likely encountering a 
molecular cloud in the west.
This conclusion is supported by our millimeter observations with
SEST and the IGPS database 
where the line emission in the $^{12}$CO and $^{13}$CO (J=1--0) 
transitions show an enhancement of brightness on the western side,
correlated with the X-ray brightness and peaking near 
$v\sim$38--40~km~s$^{-1}$. The derived distance of 10~kpc agrees
with previous estimates and strengthens the association between
3C\,397 and the molecular cloud.
Finally, we conclude that \3c\ is at the critical stage
of entering into
its radiative phase; and propose that it will eventually evolve into the
more classical mixed-morphology SNRs.

\acknowledgements
This research made use of the \chandra's X-ray Center website,
NASA's Astrophysics Data System (ADS),
the High Energy Astrophysics Science Archive Research Center
(HEASARC) operated by NASA's Goddard Space Flight Center (GSFC),
and the International Galactic Plane Survey (IGPS).
The latter is a campaign which combines data from the Dominion Radio
Astrophysical Observatory (DRAO), the Very Large Array (VLA), 
and the Australia Telescope National Facility's Compact Array (ATCA). 
The Canadian component (CGPS) is a Canadian project with
international partners, supported by a grant from the
Natural Sciences and Engineering Council of Canada (NSERC).
We thank Una Hwang for useful discussions and involvement 
in the earlier stages of this project; Steve Reynolds for
providing us with the VLA fits file; Keith Arnaud for
fixing $XSPEC$ bugs; Richard Mushotzky,
Koji Mukai and Jonathan Keohane for fruitful discussions about \cxo,
and S. Paron for his help in the CO data handling.
We are grateful to W. Miller Goss for
conducting the search for OH masers in \3c\ and for
useful discussions.

We thank the anonymous referee for a careful reading of the manuscript.
SS-H acknowledges support by an NSERC University Faculty Award,
an NSERC Discovery Grant, and a NASA LTSA grant through the
cooperative agreement NCC5-356 with Universities Space Research
Association (USRA) at GSFC.

\newpage

\begin{deluxetable}{c|ccc}
\rotate
\tablecaption{Spectral results for the eastern lobe, the central `hot spot', and the western lobe. The model used was a two-component
thermal bremsstrahlung with Gaussian lines, 
modified by interstellar absorption with column density $N_{\rm H}$.}
\tablehead{
\colhead{Parameter/Region} & \colhead{east} &\colhead{west} &\colhead{hot spot} }
\startdata
N$_{\rm H}$ (10$^{22}$~cm$^{-2}$) &     2.27 (2.24-2.31)&       2.86 (2.84-2.89)&            2.76 (2.6-3.6)\\
kT$_s$ (keV)  & 0.25 (0.239-0.251) &  0.256 (0.25-0.26) &          0.21 (0.17-0.25)\\
EM$_s$$^a$  &1.34 (1.24-1.76)       & 3.97 (3.96-4.1) &     0.9 (0.15-6.7)\\
kT$_h$  (keV) & 3.4 (3.0-4.1)&        2.87 (2.58-3.33)&      1.5 (0.93-3.3)\\
EM$_h$$^a$ ($\times$10$^{-4}$)  & 5.3 (4.5-6.2) & 13.3 (11.6-13.5)  &      1.8 (0.5--5.6)\\
\hline
E1 (Ne-Ly$\alpha$, Fe-L blend)     & 0.99 (0.95-1.0) &     1.0 (0.999-1.01)  &\\
E2 (Mg He$\alpha$)  &   1.33 (1.32-1.34)        & 1.345 (1.33-1.35) &      \\
E3 (Si He$\alpha$) &    1.825 (1.816-1.830) &   1.825 (1.815-1.828) &   1.79 (1.74-1.84)\\
E4  (S He$\alpha$) &   2.415 (2.41-2.42) &     2.418 (2.415-2.423) &   2.41 (2.38-2.44) \\
E5 (S He$\beta$?) &     2.90 (2.83-2.95) &      2.931 (2.907-2.950)&   2.87 (2.84-2.90)\\
E6 (Ar He$\alpha$)&      3.12 (3.09-3.15)&  3.1 (3.08--3.11)     & \\
E7 (Ca He$\alpha$) & & 3.826 (3.79-3.86)  & \\
E8 (Fe He$\alpha$) &    6.498 (6.48-6.51)       & 6.546 (6.53-6.56)    &  \\
\hline
Flux$_{ua}$ (0.5-8 keV), soft & 1.7$\times$10$^{-10}$     &5.30$\times$10$^{-10}$              &6.3$\times$10$^{-11}$\\
Flux$_{ua}$ (0.5-8 keV), hard & 1.7$\times$10$^{-12}$&    3.0$\times$10$^{-12}$&       3.4$\times$10$^{-13}$      \\
\hline
$\chi^2_{\nu}$ ($\nu$) &   1.75  (219) &  1.88 (265) & 0.87 (116)  \\
cts~s$^{-1}$~arcsec$^{-2}$   & 8.04$\times$10$^{-5}$&     1.33$\times$10$^{-4}$ &     2.34$\times$10$^{-4}$\\

\enddata

$^a$ the emission measure in units of
(3.02$\times$10$^{-15}$/4$\pi$$D^2$)$\int$$n_e$$n_i$$dV$
cm$^{-5}$, where $n_e$ and $n_i$ are the electron and ion
densities in cm$^{-3}$, respectively.
\end{deluxetable}

\clearpage

\begin{table}
\caption{Two-component thermal model for the hot spot.
Errors are at
the 2$\sigma$ level.}
\begin{tabular}{ccc}
\hline
Parameter & vpshock+vpshock & vpshock+mekal\\ \hline
$N_{\rm H}$ (10$^{22}$~cm$^{-2}$) & 2.94 (2.9--3.5) & 3.26  (3.0--3.4) \\
$kT_s$ (keV) & 0.19  (0.15--0.20)& 0.18 (0.15--0.21) \\
$\tau_s$ (10$^{13}$~cm$^{-3}$~s) & 3.5 & -- \\
%$norm_s$ & 0.11 &  0.41 \\
$kT_h$ (keV) & 2.7 (1.3--3.7) & 2.5 (1.4--7.6)  \\
$\tau_h$ (10$^{10}$ cm$^{-3}$~s) & 6.5 (3.6--42) & 4.1 (1.9--56)  \\
%$norm_h$ & 1.66$\times$10$^{-4}$ & 2.1$\times$10$^{-4}$ \\
S$^a$ & 5 (3--18) & 6.9 (3-15) \\ \hline
$\chi^2_{\nu}$  & 0.96 (119) & 1.0 (120) \\ \hline
\end{tabular}

$^a$ The abundances of all other elements are consistent with solar.
\end{table}

\newpage

\begin{table}[tbh]
\caption{The spectral fitting results of
the eastern and western lobes.
The model used was a two-component \vshock\ model,
modified by interstellar absorption. 
The abundances (in solar units)
of the soft component were tied to those of the hard
component (except for the Fe abundance of the hot component).
A Gaussian line near 3.1 keV has been added to account for the emission
from Argon. Errors are at the 3$\sigma$ level and were determined
after freezing the parameters of the second component to their
best fit values.}

\begin{tabular}{c|cc|}
\hline
Parameter & eastern lobe & western lobe \\ \hline
$N_{\rm H}$ (10$^{22}$~cm$^{-2}$) & 2.85 (2.7--2.9) & 3.27  (3.1--3.5) \\
$kT_s$ (keV) & 0.20  (0.195--0.215)& 0.215 (0.20--0.23) \\
$\tau_s$ (10$^{11}$~cm$^{-3}$~s) & 1.6 (1.3--2.1)  & 1.2 (0.75--1.9) \\
$norm_s$$^a$ & 4 (3--5.4) &  3.3 (2.0--5.5) \\
$kT_h$ (keV) & 1.6 (1.5--2.2) & 1.4 (1.3--1.6)  \\
$\tau_h$ (10$^{11}$ cm$^{-3}$~s) & 1.4 (1.0--2.0) & 2.7 (1.9--4.0)\\
$norm_h$$^a$ & 2.7 (2.3--3.1) $\times$10$^{-3}$ & 4.9 (4--6) $\times$10$^{-3}$ \\
\hline
O & 1.5 (1.1--1.7) & 7.4 (4.5--11.5)\\
Ne & 0.38 ($\leq$0.5) & 1.1  (0.4--2.2) \\
Mg & 0.1 ($\leq$0.2) & 0.46 (0.3--0.65)\\
Si & 0.5 (0.45--0.65)  & 1.0 (0.8--1.3)   \\
S & 1.9 (1.7--2.2) & 3.3 (2.8--4.0) \\
Ca	& 1.4 ($\leq$4) & 3.1 (1--6) \\
Fe$_h$ & 12 (9--14) & 15 (12--20) \\ \hline
$\chi^2_{\nu}$  & 2.1 (226) & 2.25 (277) \\ \hline
\end{tabular}

$^a$ the normalization in units of 
(10$^{-14}$/4$\pi$$D^2$)$\int$$n_e$$n_{\rm H}$$dV$
cm$^{-5}$, where $D$ is the distance to the source (cm),
$n_e$ and $n_{\rm H}$ are the electron and hydrogen 
densities in cm$^{-3}$, respectively.
%10**-14 / (4 pi (D_A*(1+z))**2) Int n_e n_H dV, where D_A is
%the angular size distance to the source (cm), n_e is the
%electron density (cm**-3), and n_H is the hydrogen density (cm**-3)
\end{table}

\clearpage

\begin{center}
\begin{deluxetable}{cccccccccccccc}
\tabletypesize{\scriptsize}
\tiny
\rotate
\tablecaption{Spectral Parameters of the small regions shown in Fig.~6.
The model used is a two-component $\it{vpshock}$ model. 
The errorbars are large. The 2$\sigma$ errors shown below are
determined after freezing all the other parameters to their best-fit values.
See text for details.}
\tablehead{
\colhead{Region} & \colhead{$N_{\rm H}$} &  \colhead{$kT_s$} & \colhead{$tau_s$}
 & \colhead{$kT_h$} & \colhead{$tau_h$} &
 \colhead{S}  & \colhead{Fe} & \colhead{$\chi^2_{\nu}$~($\nu$)} \\ 
\colhead{} & \colhead{(10$^{22}$~cm$^{-2}$)} & \colhead{(keV)} & \colhead{(10$^{12}$~cm$^{-3}$~s)} & (keV) & \colhead{(10$^{11}$~cm$^{-3}$~s)} & \colhead{solar} & \colhead{solar} & \colhead{} 
}
\startdata
1 & 3.03$^{+0.12}_{-0.03}$ & 0.15 & 1.05$^{+0.55}_{-0.45}$ & 1.4$^{+0.35}_{-0.25}$ & 
3.5$^{+7.5}_{-1.5}$ & 2.8$^{+2.2}_{-0.8}$ & 10($\geq$5) & 1.86 (97) \\

2 & 2.86$^{+0.04}_{-0.26}$ & 0.16 & 1.7$\geq$2 & 1.7$^{+0.4}_{-0.35}$ &
1.7$^{+2.3}_{-0.2}$ & 2.0$^{+1.6}_{-0.8}$ & 6$^{+8}_{-3}$ & 1.59 (90) \\

3 & 2.95$^{+0.15}_{-0.05}$ & 0.23 & 49($\geq$6) & 18$\geq$8 &
0.48$^{+0.22}_{-0.13}$ & 5.2$\pm$1.2 & 2.1$^{+0.7}_{-0.3}$ & 1.8 (124) \\

4 & 3.43$^{+0.11}_{-0.09}$ & 0.18 & 46($\geq$6) & 2.4$^{+0.9}_{-0.4}$ & 0.32$^{+0.18}_{-0.08}$
 &  3.3$^{+1.2}_{-0.8}$ & 1.5$^{+1.3}_{-1.4}$ & 1.63 (112) \\

5 & 3.67$^{+0.13}_{-0.12}$ & 0.19 & 2.23$^{+7.8}_{-1.7}$ & 1.53$^{+0.19}_{-0.07}$
 & 2.1$^{+0.6}_{-1.96}$ & 2.4$^{+0.3}_{-0.4}$ & 6.3$^{+1.0}_{-1.7}$ & 1.4 (189) \\

6 & 3.96$^{+0.02}_{-0.08}$ & 0.17 & 0.5$^{+0.07}_{-0.03}$ &
1.58$^{+0.12}_{-0.28}$ & 2.0$^{+3}_{-0.5}$  & 1.8$^{+0.5}_{-0.6}$ & 5.2($\geq$3.5) & 2.3 (204) \\

7 & 3.0$^{+0.21}_{-0.19}$ & 0.26 & 0.05$^{+0.03}_{-0.025}$ & 1.0$^{+0.14}_{-0.06}$ & 6$^{+12}_{-2}$ &  0.6$^{+0.3}_{-0.1}$ & 1.4$^{+0.5}_{-0.7}$ & 1.14 (129) \\

8 & 3.2$^{+0.3}_{-0.2}$ & 0.26 & 0.007$^{+0.003}_{-0.002}$ & 2.4$^{+1.0}_{-0.8}$
& 2.4$^{+9.6}_{-0.4}$ & 8($\geq$3) & 14$\pm$5 & 1.0 (70) \\

9 & 3.85$^{+0.25}_{-0.15}$ & 0.22 & 0.030$^{+0.022}_{-0.010}$ & 1.04$^{+0.11}_{-0.09}$ & 
12.5($\geq$0.5) &  1.5$^{+0.9}_{-0.4}$ & 7$^{+6}_{-3}$ & 1.19 (115) \\

10 & 3.36$^{+0.24}_{-0.06}$ & 0.23 & 26($\geq$2) & 3.4$^{+1.5}_{-0.9}$ & 0.7$^{+0.5}_{-0.2}$ & 
3.5$^{+1.5}_{-1.2}$ & 5.2$^{+2.3}_{-1.7}$ & 1.1 (107) \\

11 & 3.6$^{+0.12}_{-0.01}$ & 0.17 & 0.32$^{+0.18}_{-0.02}$ & 1.38$^{+0.32}_{-0.13}$ &
2.6$^{+2.4}_{-1.1}$ & 1.9$^{+1.6}_{-0.4}$ & 4.6$^{+2.4}_{-2.1}$ & 1.49 (139)\\

12 & 4.10$^{+0.16}_{-0.14}$ & 0.19 & 0.27$^{+0.27}_{-0.12}$ & 1.03$^{+0.21}_{-0.13}$ 
& 5.3($\geq$2.0) & 2.0$^{+0.7}_{-0.6}$ & 10$^{+8}_{-3}$ & 1.23 (118) \\

\enddata
\end{deluxetable}

\end{center}

\begin{table}[tbh]
\caption{Spectral parameters of hard point source south-east
of 3C\,397. The model is an absorbed power law with a Gaussian line.
Errors are at the 2$\sigma$ level.}
\begin{tabular}{cc}
\hline
$N_{\rm H}$ (10$^{22}$~cm$^{-2}$)   & 5.5 (4.4-6.7)\\
$\Gamma$ & 0.025 (-0.22--0.28)\\
$norm.$\tablenotemark{a} & 2.7 (1.7--4.2) $\times$10$^{-5}$ \\
$E_0$ & 6.345 (6.32--6.37)\\
$EQW$\tablenotemark{b} & 425 eV \\
$norm.$\tablenotemark{c}  & 1.08 (0.8--1.3) $\times$10$^{-5}$ \\
$\chi_{\nu}^2$ & 0.85 \\
$\nu$ & 79\\
\hline
$F_{observed}$ (1--10 keV) & 1.75$\times$10$^{-12}$ erg~cm$^{-2}$~s$^{-1}$\\
$F_{ua}$ (1--10 keV)\tablenotemark{d} & 2.1$\times$10$^{-12}$ erg~cm$^{-2}$~s$^{-1}$\\
\vspace{-2cm}
\tablenotetext{a}{Normalization of the power law model
in units of photons keV$^{-1}$ cm$^{-2}$ s$^{-1}$ at 1 keV.} \\
\tablenotetext{b}{Equivalent Width.}\\
%For the thermal bremsstrahlung model, the normalization unit is
%(3.02$\times$10$^{-15}$/4$\pi$$D^2$)$\int$$n_e$$n_i$$dV$
%cm$^{-5}$, where $n_e$ and $n_i$ are the electron and ion
%densities in cm$^{-3}$, respectively. 
%%
\tablenotetext{c}{Total photons~cm$^{-2}$ s$^{-1}$ in the line.}\\
\tablenotetext{d}{Unabsorbed flux.}
\end{tabular}
\end{table}

\newpage

\begin{table}[tbh]
\caption{Millimeter Line Flux density in \3c. See \S5 for details.}
\begin{tabular}{ccccc} 
\hline
Location & Antenna Temp. & Antenna Temp. & Line Flux Density  & Line
Flux Density\\
 & $^{12}$CO J=1-0 & $^{12}$CO J=2-1 & $^{12}$CO J=1-0 & $^{12}$CO J=2-1 \\
   & (K) & (K)&  (Jy) &  (Jy) \\ \hline 
Center &  3.7 & 2.3 &  1.43 & 1.24\\
North  & 2.81 & 1.56 & 1.08 & 0.84\\
South & 3.82 & 2.08 & 1.47 & 1.12\\
East & 2.49 & 1.5 & 0.96 & 0.52\\
West & 4.79 & 2.72 & 1.85 & 1.00\\
\hline
\end{tabular}
\end{table}

\clearpage
\newpage

\begin{figure}[tbh]
%\centerline{\plotone{../f1.eps}}
\caption{The energy image of \3c. Red, green, and blue correspond to the 0.5--1.5 keV,
1.5--2.5 keV, and 2.5--8 keV ranges, respectively. The individual images
have been adaptively smoothed using a Gaussian kernel and an FFT
algorithm, with a minimum (maximum) smoothing scale of 2 (4) pixels,
and with a minimal (maximal) signal to noise ratio of 3 (5). }
\end{figure}

\begin{figure}[h]
%\epsscale{1.2}
%\centerline{\plottwo{../f2a.eps}{../f2b.eps}
%}
%\epsscale{1.0}
\caption{The \chandra\ S3 smoothed images of the central X-ray spot (left)
and the hard point-source southeast of \3c\ (right). 
Both images are shown on the same scale.}
\end{figure}

\begin{figure}[h]
%\centerline{\plotone{f3.eps}}
\caption{The \chandra\ image in the 0.5--2.4 keV band,
with the \rosat\ HRI contours overlayed. The \chandra\ image has been
smoothed with the ciao tool $csmooth$ using a tophat filter with 2$^{\prime\prime}$ (5$^{\prime\prime}$)
minimum (maximum) scale and a minimum significance of 3$\sigma$. 
The HRI image
has been smoothed with a Gaussian with sigma=4$^{\prime\prime}$ to show the enhanced emission
from the central spot centered at $\alpha$=19$^h$ 07$^m$ 35$^s$.13, $\delta$=07$^o$ 08$^{\prime}$ 28$^{\prime\prime}$.76
(J2000). The arrows point to the soft and hard point sources discussed in \S3 and \S4.4.}
\end{figure}

\begin{figure}[h]
%\epsscale{0.75}
%\centerline{\plotone{../f4.eps}}
%\epsscale{1.0}
\caption{The \chandra\ 0.3--10 keV image with the VLA L-band 
contours overlayed.} 
\end{figure}

\begin{figure}[h]
%\epsscale{0.85}
%\centerline{\plotone{../f5.eps}
%}
%\epsscale{1.0}
\caption{The \chandra\ 4 to 6 keV image (color) with
the radio VLA L-band contours overlayed.}
\end{figure}

\begin{figure}[h]
%\epsscale{1.0}
%\centerline{\plotone{f6.eps}
%}
%\epsscale{1.0}
\caption{The regions selected for spatially resolved spectroscopy
(\S4).}
\end{figure}

\begin{figure}[h]
%\epsscale{1.0}
%\centerline{\plotone{../f7.eps}
%}
%\epsscale{1.0}
\caption{The distribution of the column density, $N_{\rm H}$, 
expressed in units of 10$^{22}$~cm$^{-2}$.
$N_{\rm H}$ was derived using the two-component \vshock\ models
summarized in Table~4.
Note that the image contrast has been suppressed
to display the numbers. }
\end{figure}

\begin{figure}[h]
%\epsscale{1.2}
%\centerline{\plottwo{../f8a.eps}{../f8b.eps}
%}
%\epsscale{1.0}
\caption{The distribution of the soft component's temperature
(expressed in keV, left) and ionization timescale
 (in units of 10$^{12}$~cm$^{-3}$~s$^{-1}$, right). 
See Table~4 and \S4.2 for details.}
% The image contrast has been wiped out to show the numbers.}
\end{figure}

\begin{figure}[h]
%\epsscale{1.2}
%\centerline{\plottwo{../f9a.eps}{../f9b.eps}
%}
%\epsscale{1.0}
\caption{The distribution of the hard component's
temperature (expressed in keV, left) and ionization timescale
 (in units of 10$^{11}$~cm$^{-3}$~s$^{-1}$, right).
See Table~4 and \S4.2 for details.}
% The image contrast has been wiped out to show the numbers.}
\end{figure}

\begin{figure}[h]
\epsscale{0.8}
\centerline{\plotone{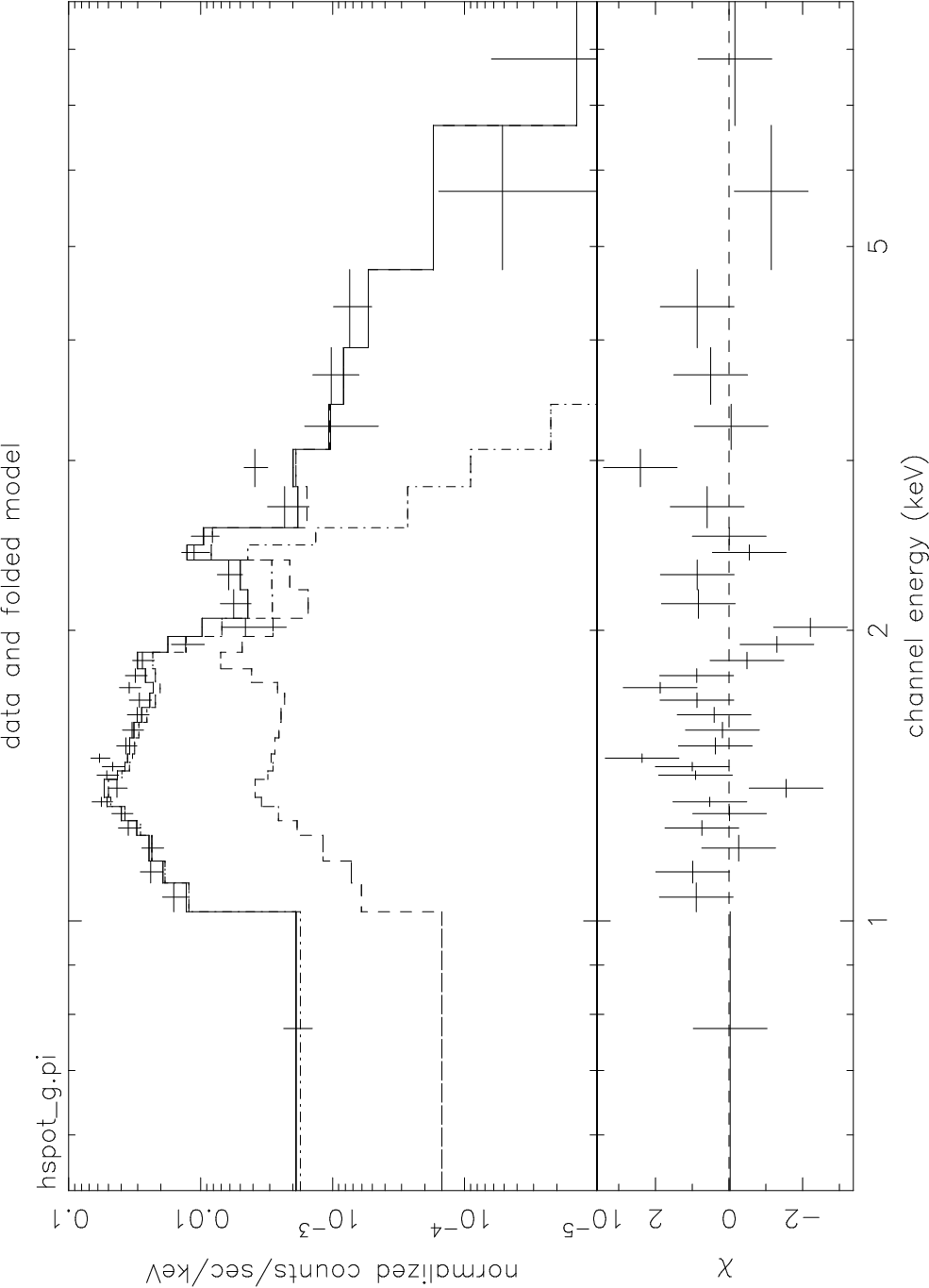}
}
\epsscale{0.8}
\caption{
The data (crosses) and the fitted model for
the central 'hot spot' in \3c.
The model used  is a two-component \vshock\ model.
The components of the model are shown as dashed lines.
The bottom panel displays the residuals in units of sigmas.
The parameters of the fits are summarized in Table~2.}
\end{figure}

\begin{figure}[h]
\epsscale{1.0}
\centerline{\plottwo{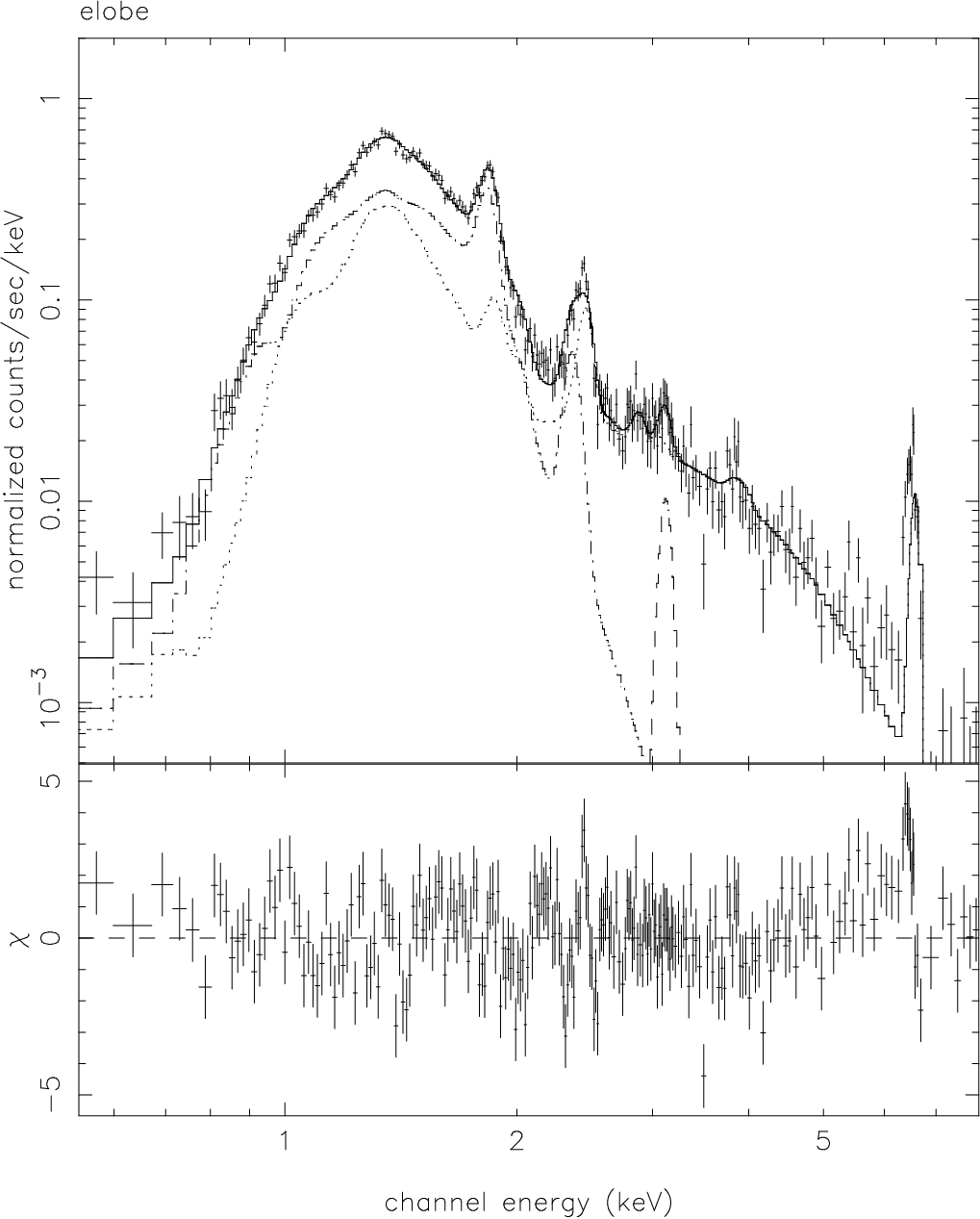}{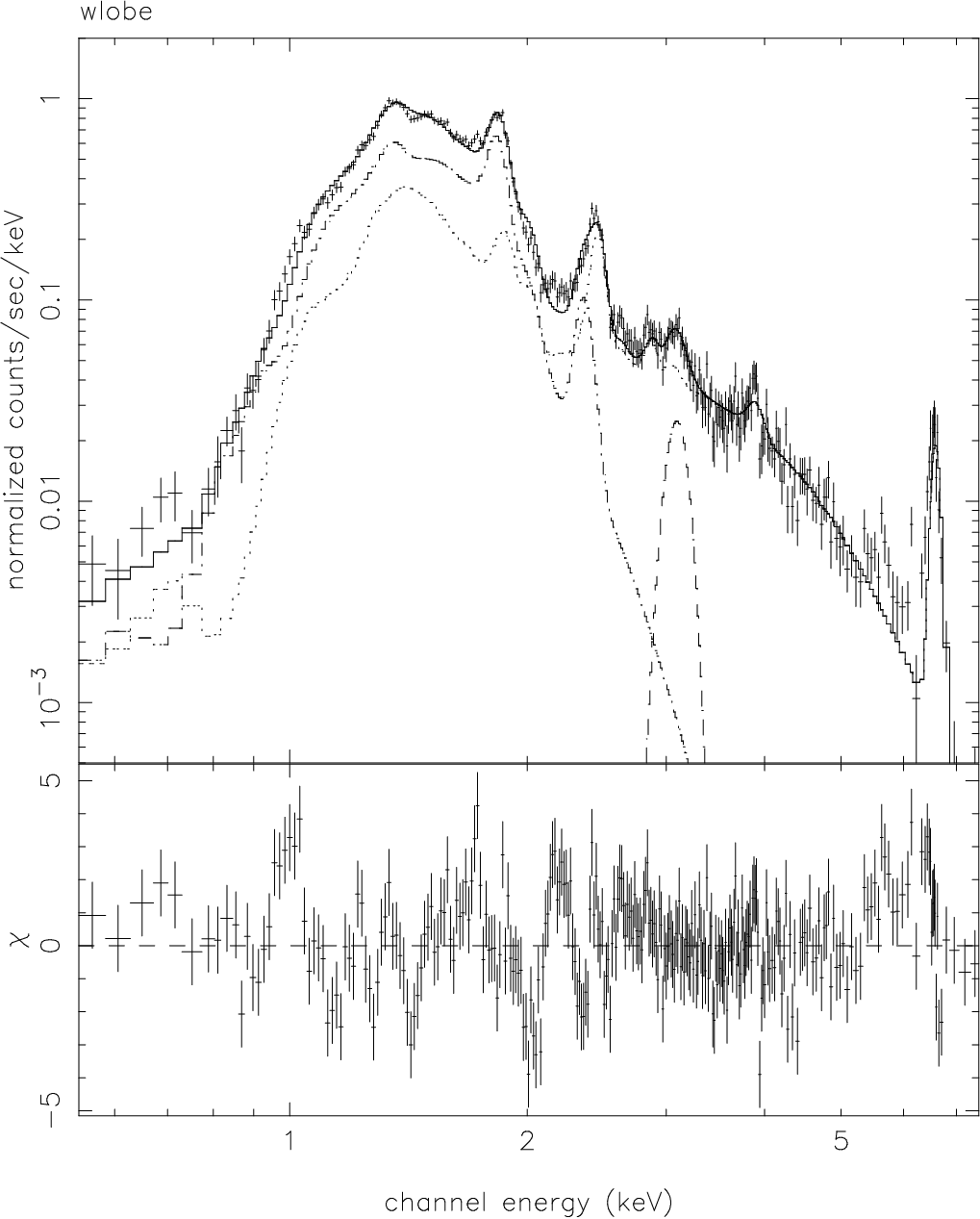}}
\epsscale{1.0}
\caption{The data (crosses) and the fitted model for
the eastern (left) and western (right) lobes of \3c.
The model used  is a two-component \vshock\ model.
 A Gaussian line
has been added to account for the line emission from Argon. 
The components of the model are shown as dashed lines.
The bottom panel displays the residuals in units of sigmas.
The parameters of the fits
are summarized in Table~3. Both plots are shown on the same y-scale.}
\end{figure}

\begin{figure}[h]
\epsscale{1.0}
\centerline{\plottwo{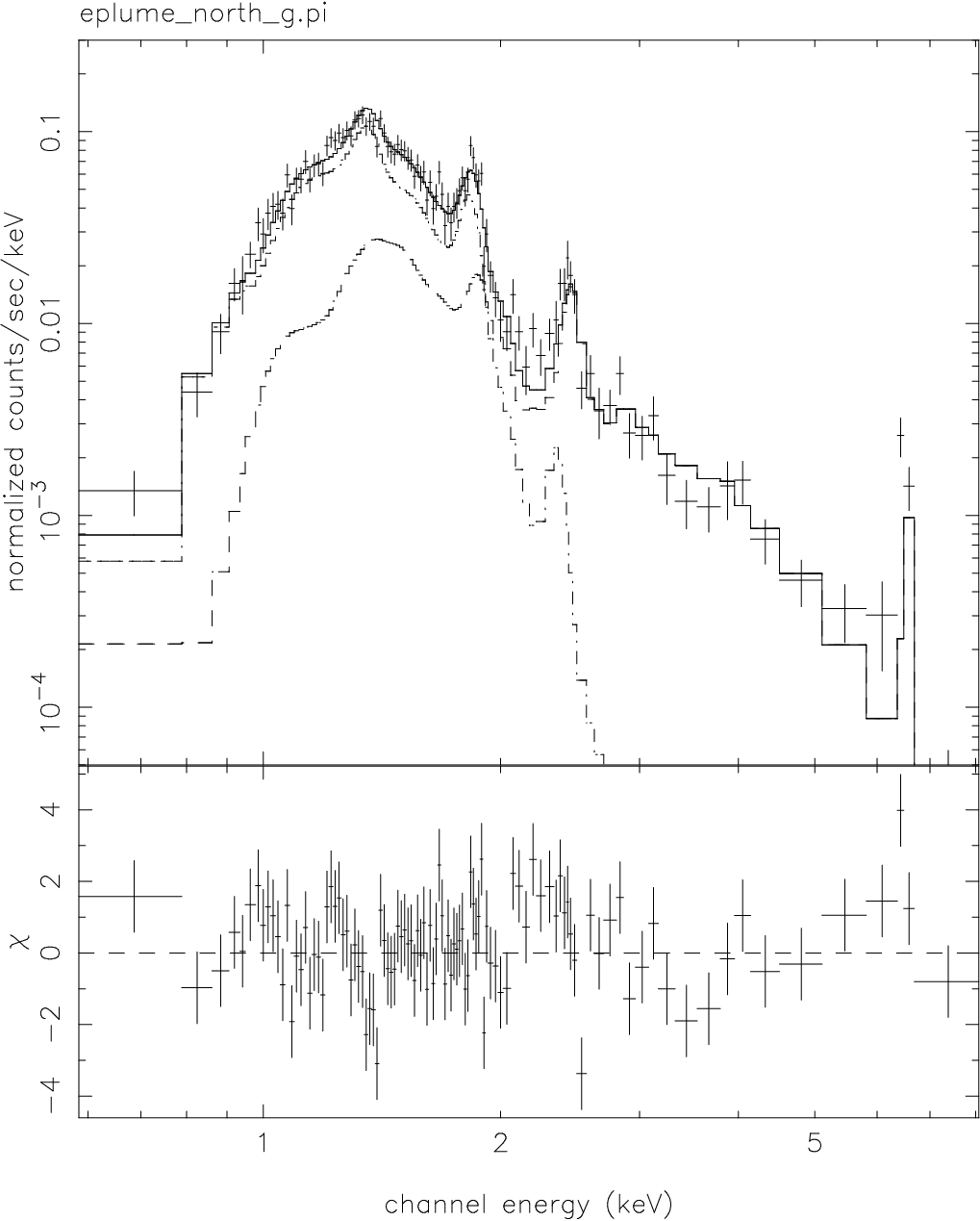}{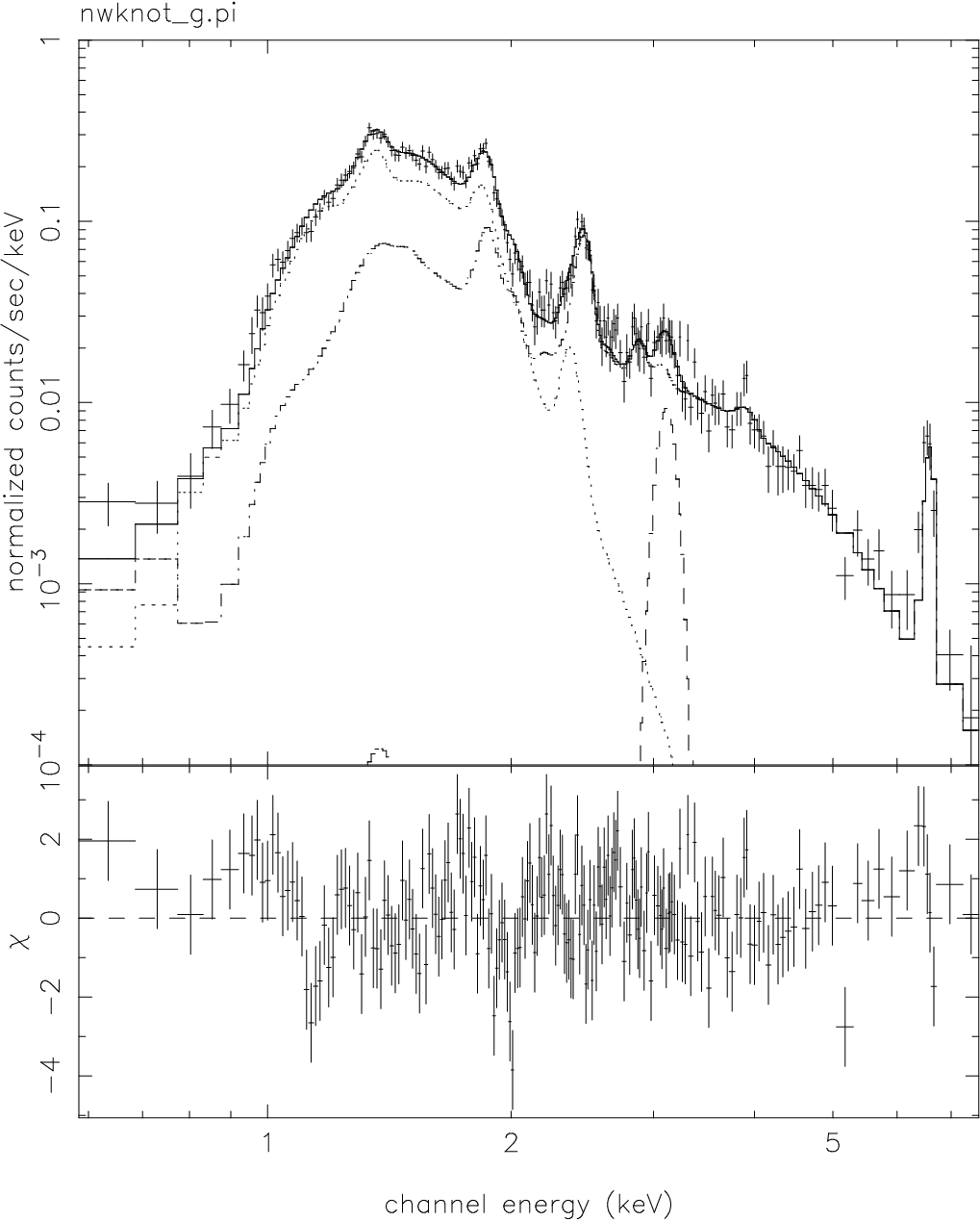}}
\epsscale{1.0}
\caption{The soft north-eastern plume (region 1 located in the eastern
lobe) and the north-western knot (region 5 located
in the western lobe) fitted with a two-component \vshock\ model.
A Gaussian line
has been added to account for the line emission from Argon.
The parameters of the model are summarized in Table~4. See \S4 for
details.}
\end{figure}

\begin{figure}[h]
\epsscale{1.0}
\centerline{\plottwo{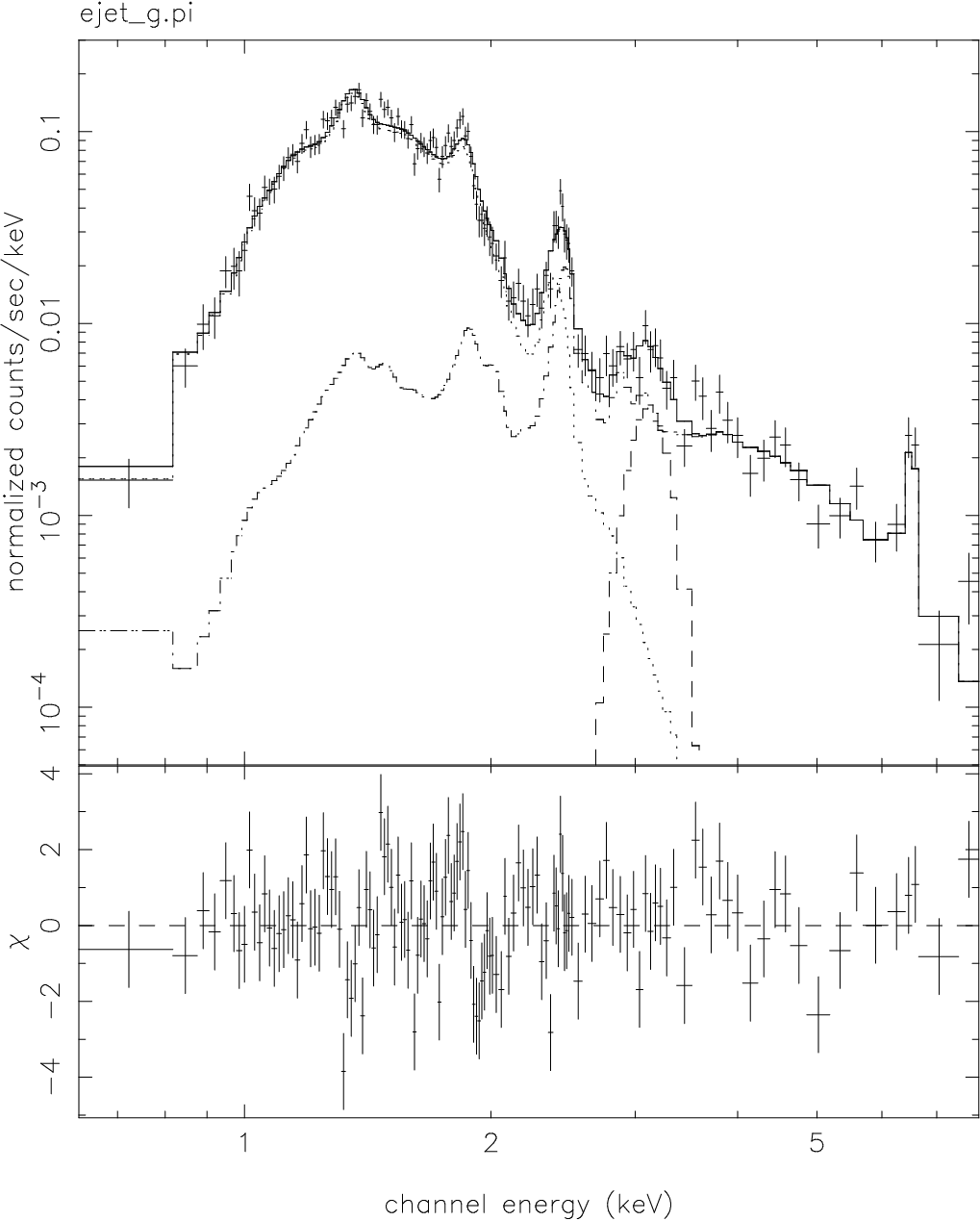}{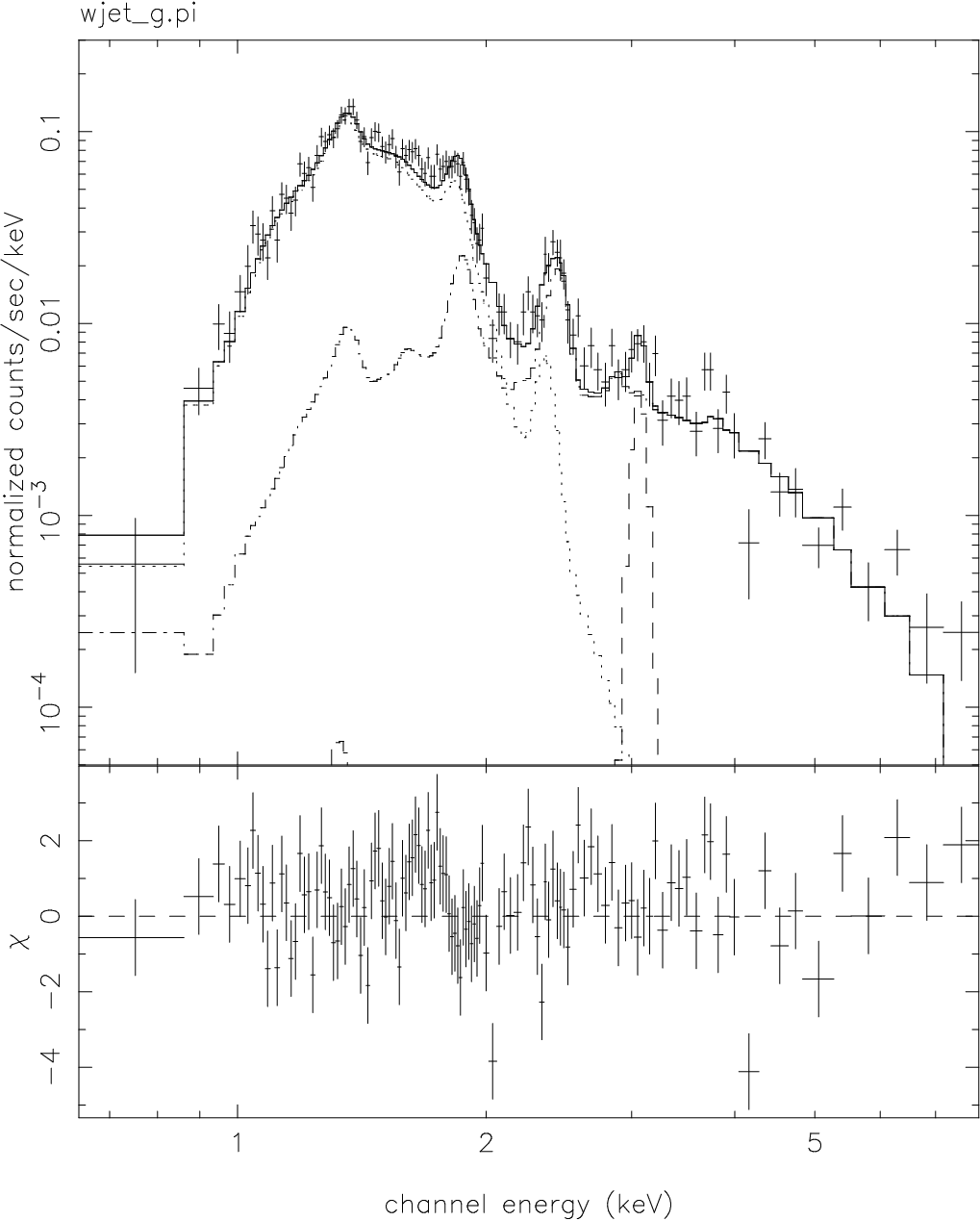}}
\epsscale{1.0}
\caption{The eastern (left) and western (right) jet-like
features (regions 3 and 4) fitted with a two-component \vshock\ model. 
A Gaussian line
has been added to account for the line emission from Argon. The parameters of the fits
are summarized in Table~4. Both plots are shown on the same y-scale.}
\end{figure}

\begin{figure}[h]
\epsscale{1.0}
\centerline{\plottwo{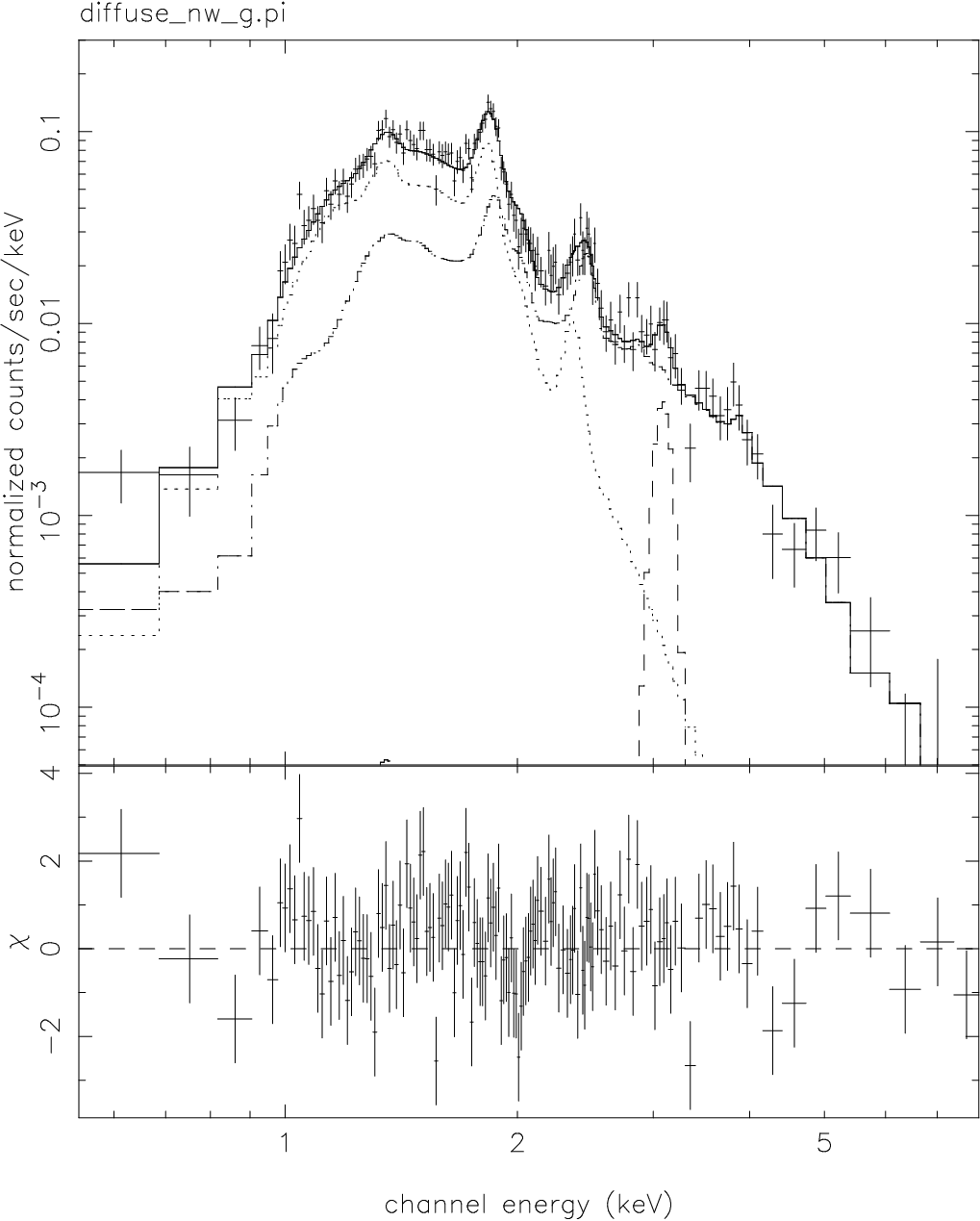}{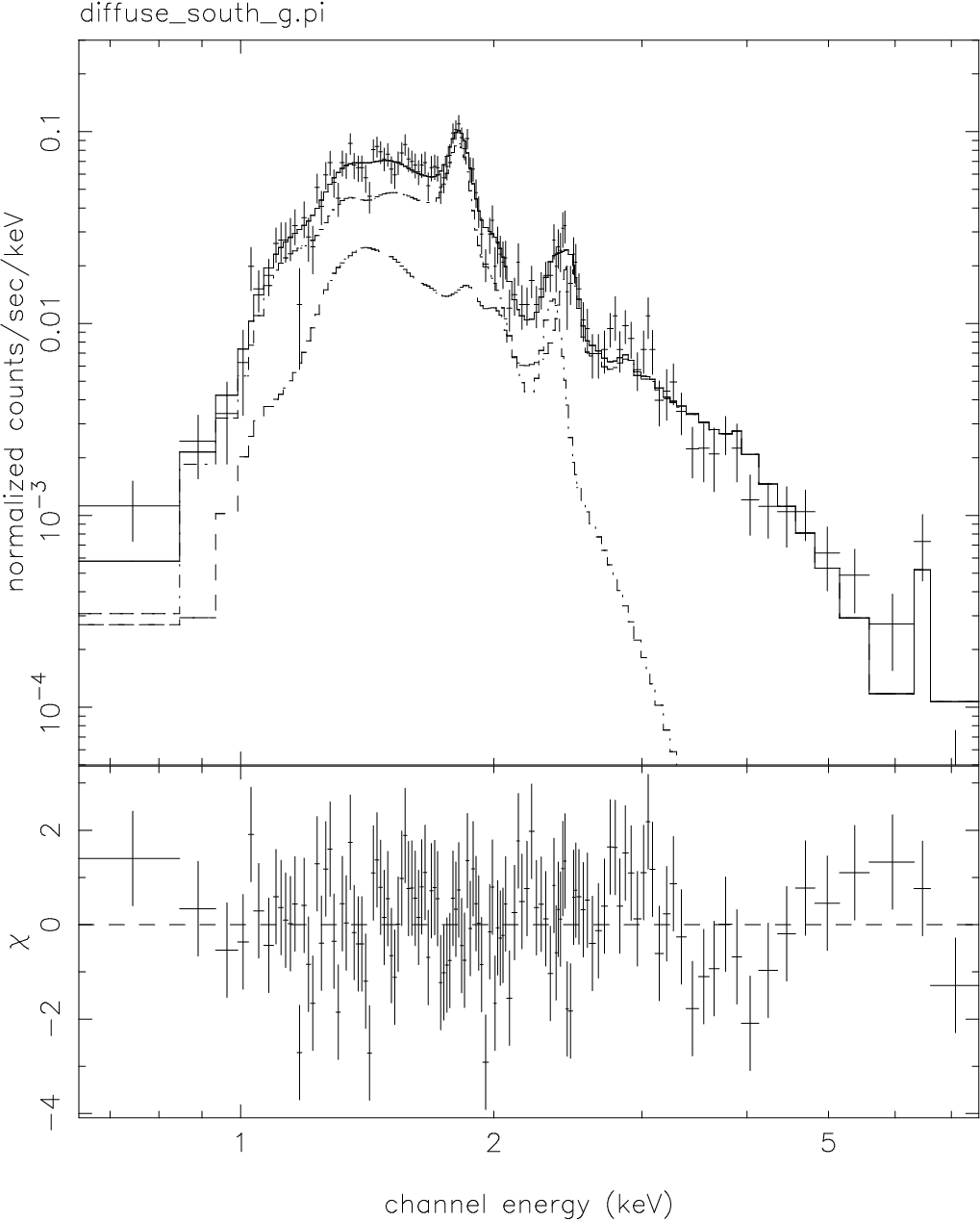}}
\epsscale{1.0}
\caption{The spectra of the low-surface brightness regions northwest
and southwest of the hot spot (regions 7 and 9),
fitted with a two-component \vshock\ model. A Gaussian line
has been added to account for the line emission from Argon. The parameters of the fits
are summarized in Table~4.}
\end{figure}

\begin{figure}[h]
\epsscale{1.0}
\centerline{\plottwo{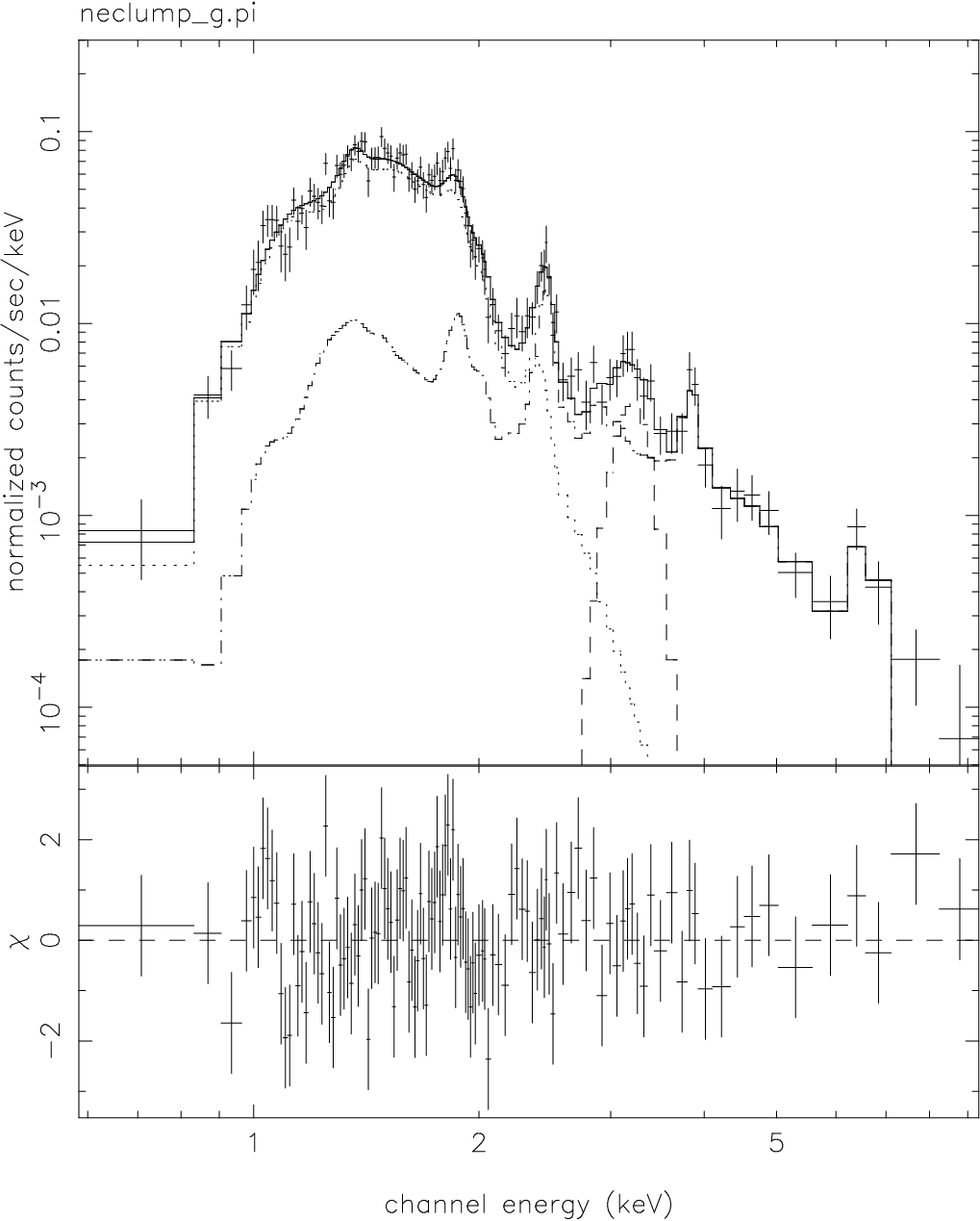}{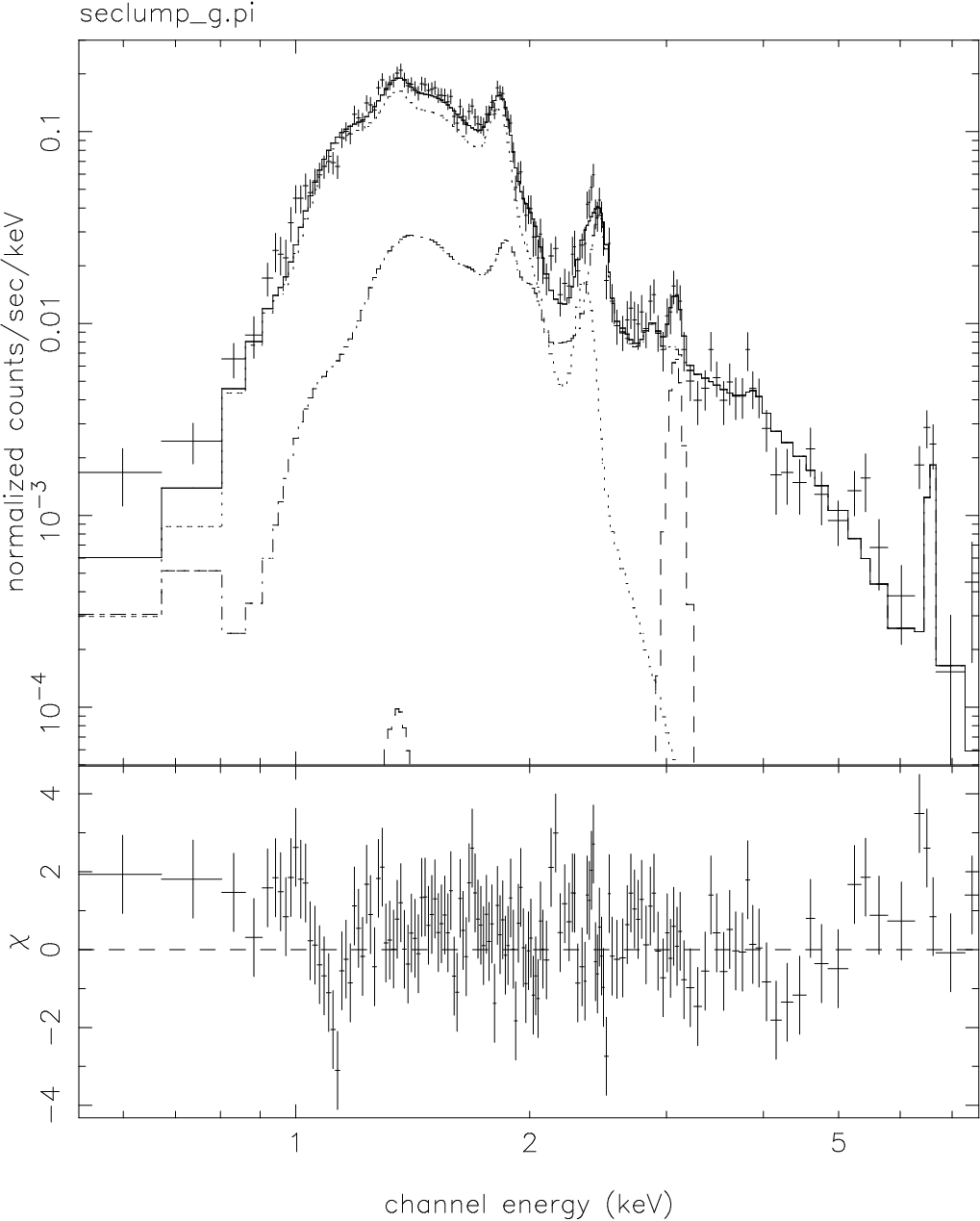}}
\epsscale{1.0}
\caption{The spectra of the northeastern and southeastern clumps
(regions 10 and 11),
 fitted with a two-component \vshock. A Gaussian line
has been added to account for the line emission from Argon.
 The parameters of the fits
are summarized in Table~4.}
\end{figure}

\begin{figure}[h]
\epsscale{0.7}
\centerline{\plotone{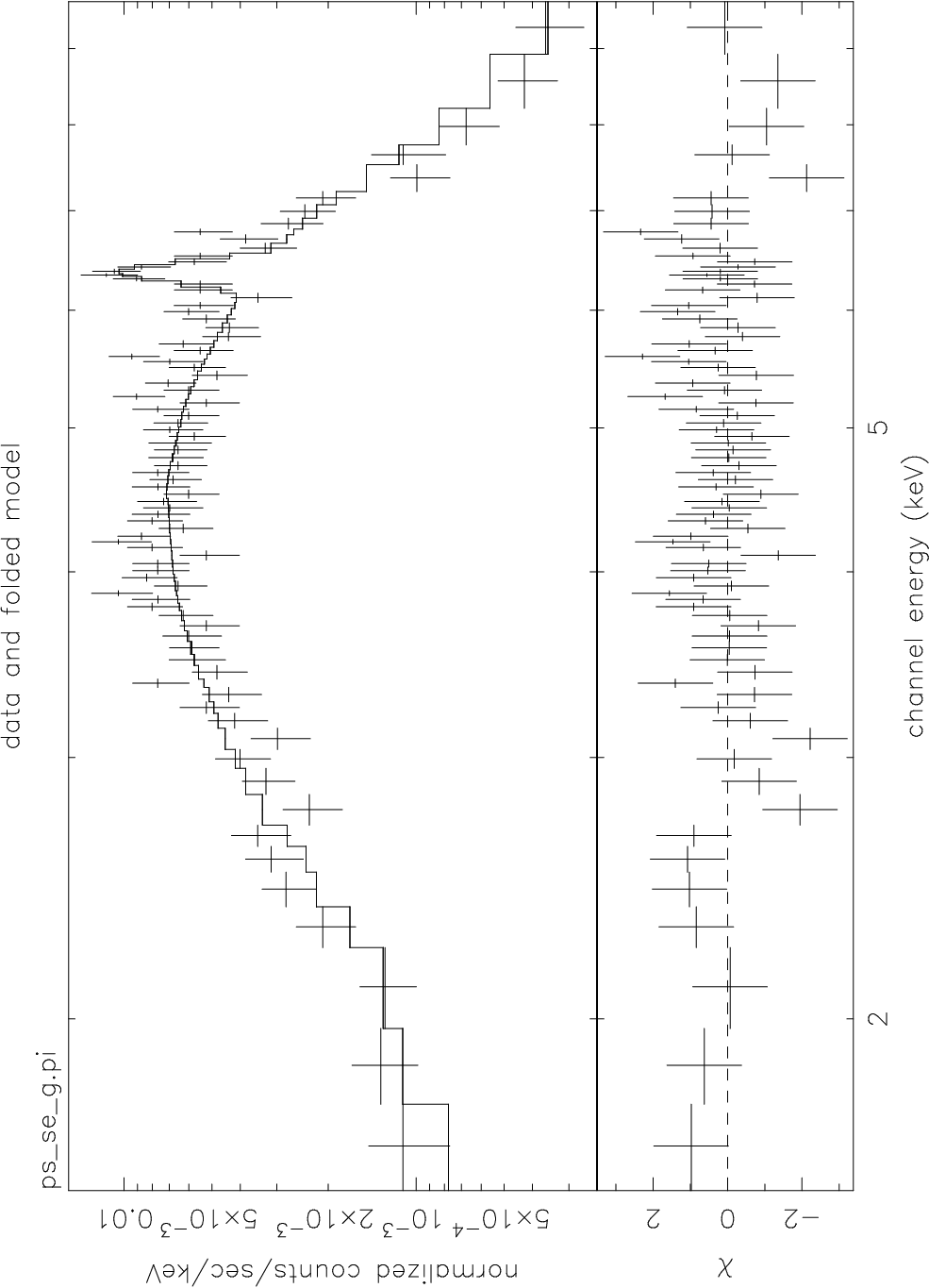}}
\epsscale{1.0}
\caption{The spectrum and fitted model of \cxo. See Table~5 for a summary
of the spectral parameters. }
\end{figure}
\newpage

\begin{figure}[h]
\epsscale{0.8}
\centerline{\plottwo{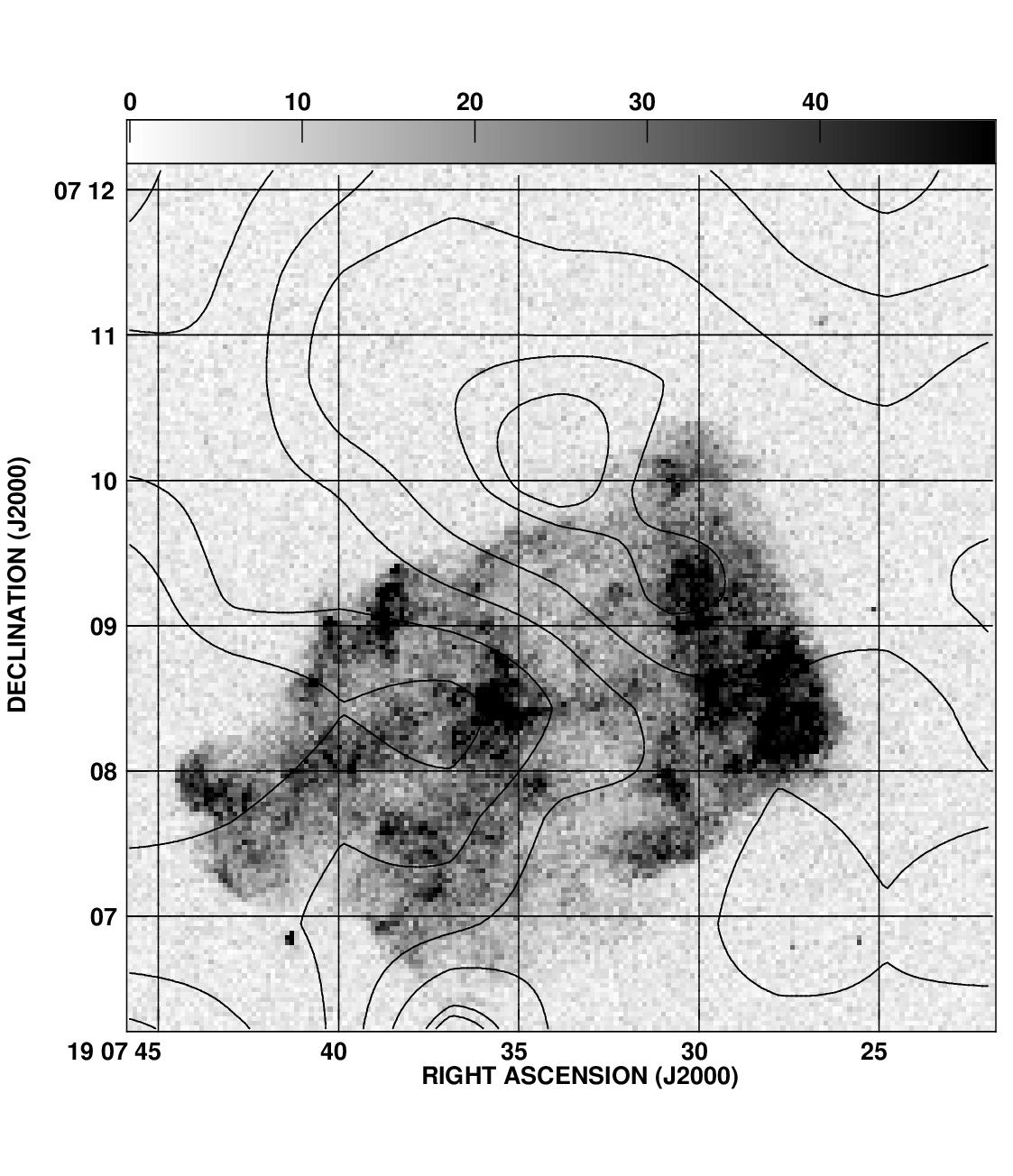}{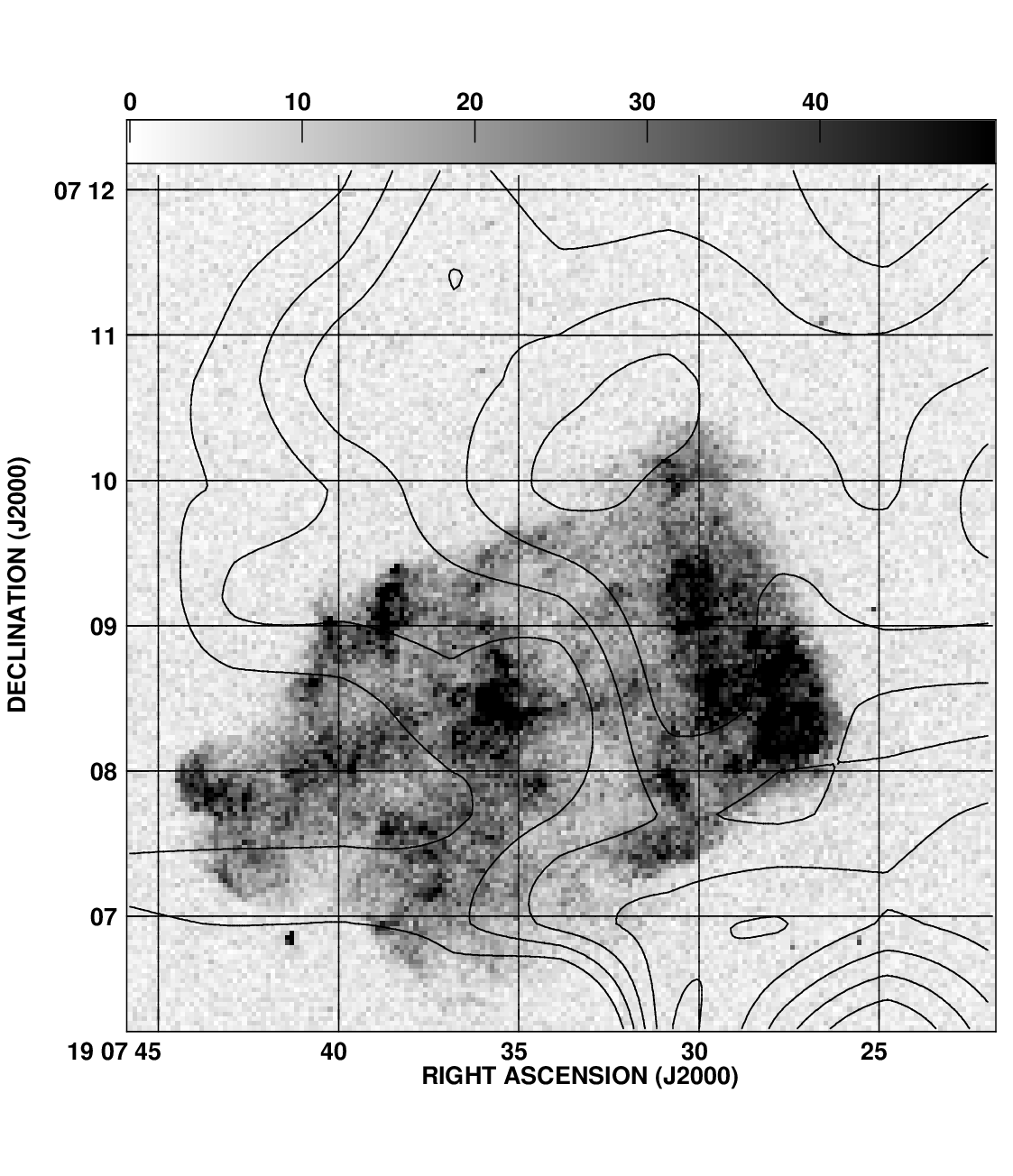}}
\epsscale{1.0}
\caption{The $^{12}$CO J=1--0 and the $^{12}$CO J=2--1 contours
 obtained with SEST with the \chandra\ image overlaid. The SEST
map was produced using a 9$\times$9~pixel map, with 45$^{\prime\prime}$
pixel to pixel separation and a mean velocity of 40~km~s$^{-1}$. The
plotted contours correspond to antenna temperatures and are: 30, 34, 38, 42,
46, 50 and 51 K for the $^{12}$CO J=1--0 map;
and  21, 23, 25, 27, 29, 31 and  33 K for the $^{12}$CO J=2--1 map.}
\end{figure}

\begin{figure}[h]
%\epsscale{0.8}
%\centerline{\plotone{f18.ps}}
%\epsscale{1.0}
\caption{The $^{13}$CO J=1--0 images in the environs
of 3C\,397, as taken form the International Galactic Plane Survey (IGPS)
for several velocity channels around $v\sim$40 km~s$^{-1}$. The gas distribution
is displayed in greyscale.
% in galactic coordinates. 
The contours represent the \chandra\ X-ray image of 3C\,397 
(smoothed with a 3$^{\prime\prime}$ gaussian). }
% this representation West is up and North to the left.}
\end{figure}

\begin{figure}[h]
%\epsscale{1.0}
%\centerline{\plotone{f19.ps}}
%\epsscale{0.75}
\caption{The $^{13}$CO J=1--0 image of \3c\ obtained with
the IGPS and  integrated between 35.4 and
41.3 km~s$^{-1}$. The CO distribution is shown in greyscale
and the contours outline the \chandra\ X-ray emission
from \3c.}
% In this representation West is up and North is to the left.}
\end{figure}

\end{document}